# Zero-index and Hyperbolic Metacavities: Fundamentals and Applications

Zhiwei Guo, Haitao Jiang, and Hong Chen[*]

*Key Laboratory of Advanced Micro-structure Materials, MOE, School of Physics Science and Engineering,*

*Tongji University, Shanghai 200092, China*

**OUTLINE**

---

**\* Corresponding author:** hongchen@tongji.edu.cn.

# ABSTRACT

As a basic building block, optical resonant cavities (ORCs) are widely used in light manipulation; they can confine electromagnetic waves and improve the interaction between light and matter, which also plays an important role in cavity quantum electrodynamics, nonlinear optics and quantum optics. Especially in recent years, the rise of metamaterials, artificial materials composed of subwavelength unit cells, greatly enriches the design and function of ORCs. Here, we review zero-index and hyperbolic metamaterials for constructing the novel ORCs. Firstly, this paper introduces the classification and implementation of zero-index and hyperbolic metamaterials. Secondly, the distinctive properties of zero-index and hyperbolic cavities are summarized, including the geometry-invariance, homogeneous/inhomogeneous field distribution, and the topological protection (anomalous scaling law, size independence, continuum of high-order modes, and dispersionless modes) for the zero-index (hyperbolic) metacavities. Finally, the paper introduces some typical applications of zero-index and hyperbolic metacavities, and prospects the research of metacavities.

## I. INTRODUCTION

Optical devices fundamentally count on the interactions between light and the matter [1]. As one of the most fundamental optical elements, optical resonant cavities (ORCs) design is crucial for effectively enhancing the light-matter interaction [2, 3]. The traditional Fabry-Pérot (FP) ORC is composed of two optical mirrors separated by a certain distance, in which light will circulate multiple times, as shown in Fig. 1(a) [4]. The basic physical principle of ORCs is the interference of light, which leads to the formation of standing wave in the cavity. Due to the important role of high-quality factor ($Q$) and ultra-small ORCs in optics, such as low threshold laser [5, 6], high resolution sensor [7-9] and so on, it has attracted people's wide attention. In general, in order to realize the ORCs with good electromagnetic (EM) wave confinement, a cavity with high-$Q$ is needed. However, the high-$Q$ ORCs usually need to increase the number of round trips, such as the optical whispering gallery mode (WGM) cavity in Fig. 1(b) [10, 11]. This leads to the increase of cavity volume ($V$), which is not conducive to the miniaturization of the ORCs. On the other hand, although the surface plasmon polaritons (SPPs) can be used to design miniaturized ORCs, the $Q$ of cavities is significantly reduced due to the large inherent loss of metal [12]. Therefore, it is difficult to achieve high-$Q$ cavity mode in small ORCs and it is very meaningful to evaluate the $Q/V$ relation in practical applications. Especially, $Q$ and $V$ correspond to the lifetime and energy density of cavity mode, respectively. With the emergence of photonic crystals (PCs), researchers have proposed PC microcavity [13, 14]. The research of PC cavities is mainly focused on one-dimensional (1D) and two-dimensional (2D) PCs, as shown in Figs. 1(c) and 1(d), respectively. Compared with the traditional ORCs in Fig. 1(a), PC microcavity based on photonic band gap has the advantages of high-$Q$ and small-$V$ (the geometric dimension is the same order of magnitude as the wavelength of light). Usually, PC cavities are realized by breaking the symmetry of structure, such as introducing point defects in PCs. As a result, light can be trapped in the position of defect, and cannot propagate to the surrounding, thus forming a PC cavity. So far, PC cavities have been used in low-threshold lasers, high-sensitivity sensors, filters and other optical devices [15]. With the development of science and technology, the traditional FP and even the PC cavities become more and more difficult to meet people's requirements for multi-function optical devices. For example, the field intensity distribution is inhomogeneous because the cavity mode exists in the form of standing wave. For the cavity quantum electrodynamics (CQED), in order to achieve strong coupling between the single atoms and the electromagnetic mode of an ORC, it is necessary to put the atoms exactly in the strongest position of the field. However, due to the small size of quantum dots, it is very difficult to accurately place the quantum dots at the antinode of the cavity mode [16-18]. In addition, with the development of on-chip miniaturized photonic devices, the realization of subwavelength optical mode localization becomes a very important scientific problem.

Recent developments in the photonics and metamaterials are opening alternative avenues for ORCs design. Metamaterials are artificial materials with ingeniously constructed, and their properties come from their structure and carefully designed meta-atoms rather than from the properties of their constitutive materials [19-22]. Especially, the period and the size of meta-atom are much smaller than the wavelength of EM waves, thus metamaterials can be described by the local EM parameters (permittivity $\varepsilon$ and permeability $\mu$) based on the effective medium theory (EMT). By choosing the appropriate meta-atom, values and symbols of $\varepsilon$ and $\mu$ can be adjusted flexibly. So far, a variety of metamaterials with very unusual electromagnetic properties and functions have been proposed, such as double-negative metamateirals (DNG, $\varepsilon < 0$, $\mu < 0$) [23, 24], μ-negative metamateirals (MNG, $\varepsilon > 0$, $\mu < 0$) [25], zero-index metamateirals (ZIM $n = \sqrt{\varepsilon}\sqrt{\mu} \approx 0$) [26-30] and gradient metamaterials [31, 32]. In addition, indefinite metamateirals with anisotropic $\varepsilon$ or $\mu$ are proposed [33], in which the hyperbolic metamaterials (HMMs), whose the principal components of electric or magnetic tensor have opposite signs, have attracted considerable interest because of their special open dispersion curves [34-40]. Metamaterials with special EM parameters can achieve many phenomenon and functions beyond the conventional materials, such as negative refraction [23], cloaking [31, 32], super-resolution imaging [41] and optical black hole [42]. The novel EM properties produced by metamaterials have good control for the transmission and radiation of EM waves. On the basis,, metamaterials [43-51] and metasurfaces [52-60] are further proposed to construct special ORCs in recent years, as shown in Fig. 1(e). Metacavities can break through many physical limitations of traditional ORCs and realize many novel functions and applications, such as the topological cavities [61-67] and vortex beams [68-73]. In this review, we mainly introduce the basic physical properties and some potential applications of zero-index metacavities [74-89] and hyperbolic metacavities [90-112].

For traditional FP and PC cavities, destructive and constructive interferences appear in different regions due to the determined optical path difference of EM waves, namely local interference. As mentioned above, due to the limitation of interference mechanism, it is difficult to achieve strong coupling between quantum dots and a traditional ORC. However, the EM field in ZIM is uniform, which can be used to achieve the global coherent effect, that is, nonlocal interference phenomenon. Therefore, ZIMs provide a good platform to study the CQED without precise positioning [76, 77]. On the other hand, because of the limitations of the standing-wave formation conditions for FP-type resonance, the miniaturization of ORCs formed using traditional materials is difficult. However, due to the special dispersion curve of HMM, it can also be used to observe the anomalous scaling law and realize size-independent ORCs. Therefore, HMMs provide a good solution for the technologies that require the miniaturization of subwavelength ORCs without reducing the value of $Q$ [93, 94]. Zero-index and hyperbolic metacavities greatly deepen our understanding of ORCs, reveal many physical phenomena rarely observed in nature, and offer applications for controlling light, including switching, unidirectional transmission, lasers, sensors, filters, and so on.

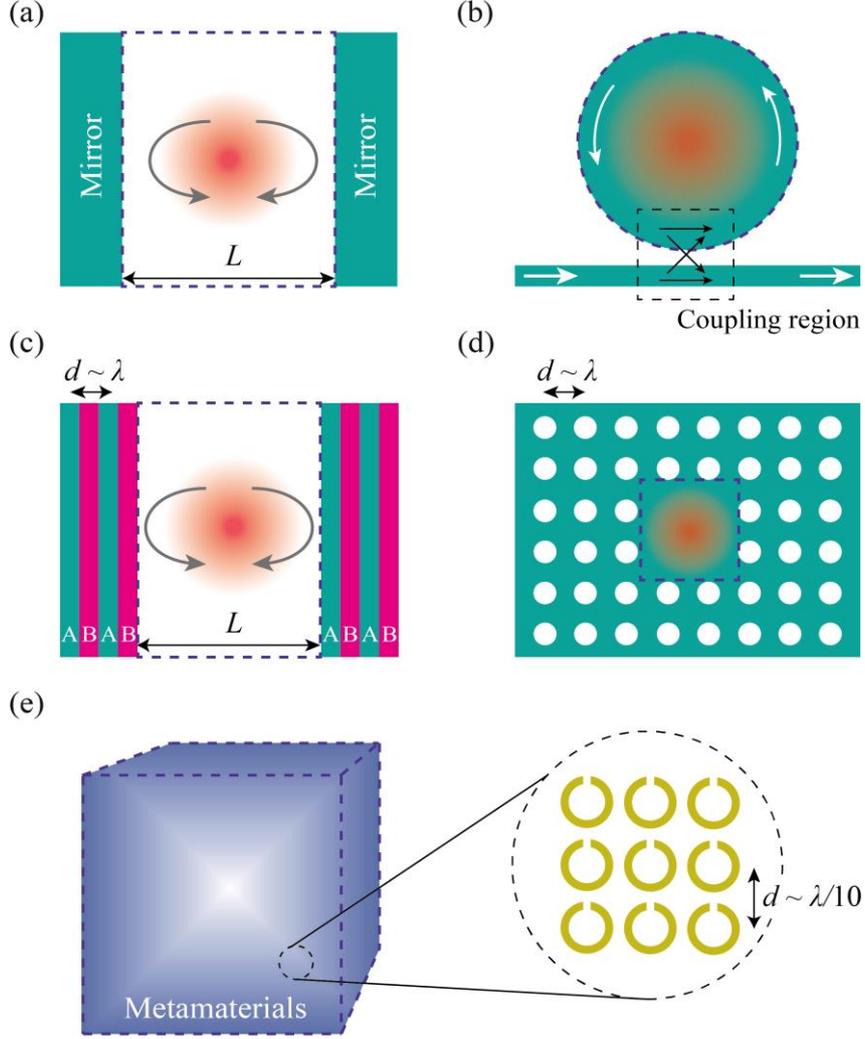

FIG. 1. **Schematics of some typical optical cavities**. (a) The conventional FP cavity composed of two optical mirrors. (b) Microdisk/Microsphere whispering gallery mode cavity. (c) 1D PC microcavity. (d) 2D PC microcavity. (e) Metacavities, whose cavity media corresponds to the artificial materials composed of subwavelength unit cells.

## II. ZERO-INDEX METACAVITES

As mentioned in the introductory part, $\varepsilon$ and/or $\mu$ are near zero in ZIMs, which can realize various EM responses and provide new situations to revisit some fundamental phenomena in wave propagations [113]. When both $\varepsilon$ and $\mu$ are near zero, the ZIM is also called an $\varepsilon$-$\mu$-near-zero (EMNZ) media. Besides, if only $\varepsilon$ or only $\mu$ is near zero, the ZIM is also called an $\varepsilon$-near-zero (ENZ) or a $\mu$-near-zero (MNZ) media. The corresponding EM parameter space of ENZ, MNZ, and EMNZ media are marked respectively by red, blue, and purple in Fig. 2(a). Because the refractive index of ZIMs tends to zero, the wavelength of EM waves tends to infinity. Thus, the phases of EM waves in ZIMs are the same everywhere [114] and the wavelength is greatly stretched [115, 116]. As a result, the light-matter interaction in ZIMs can be greatly affected [117]. Especially, because the iso-frequency contour (IFC) of ZIM is approximately a point, it can only support normal incidence EM waves and the oblique incidence EM wave will be reflected, thus it can be used to design high directivity emission [118-120]. Besides, diversified transmission characteristics of EM wave have also been verified in ZIMs [121-124]. For example, EM waves can pass through a narrow ZIM waveguide of arbitrary shape almost without blocking, which can be used to realizing the EM tunneling [125, 126]. It is well known that semiconductor doping can greatly improve the electrical properties of semiconductor devices. Inspired by this idea, doping ZIM can significantly improve its ability of EM manipulation [127], such as general impedance matching [127-130], extraordinarily large optical cross section [131], cloak [132, 133], total transmission and super reflection [134-136].

So far, most of the researches are focused on isotropic ZIMs. But another kind of special ZIM, anisotropic zero-index metamateiral (AZIM), has attracted more and more attention because of its special EM properties. In AZIM, because only one tensor element in the $\varepsilon$ or $\mu$ tensor is near zero. In this case, the IFC of AZIM will have many forms, which can be either flat elliptic or hyperbolic [137], as shown in in Fig. 2(d). Therefore, compared with isotropic ZIMs, AZIMs have more abundant EM response characteristics. For example, Luo *et al*. proposed a near perfect bending waveguides based on AZIMs [138, 139]. Interestingly, it has been found that sub-wavelength flux manipulation can be achieved in AZIMs by using the scattered evanescent waves [140], which was confirmed by experiments based on the circuit-based AZIM [141]. In general, ZIM show abundant EM response characteristics when the EM wave interacts with it because of its special EM parameters, which provides a very good platform for realizing more novel physical phenomena and important applications.

## A. Realization ways

### 1. Two-dimensional photonic crystal ZIMs

The simplest way to achieve ZIMs is to use the natural dispersion characteristics of materials. For metal materials, the permittivity can be described by a simple Drude model $\varepsilon = \varepsilon_{\infty} - \omega_p^2 / [\omega(\omega + i\Gamma)]$, where $\varepsilon_{\infty}$ is the high-frequency permittivity, $\omega$ is the angular frequency, $\Gamma$ denotes the loss of the material and $\omega_p$ is the plasma frequency. Generally, the real part of the $\varepsilon$ is close to zero when $\omega \to \omega_p$ and $\varepsilon_\infty \approx 1$. Although this method is suitable for the fabrication of isotropic ENZ media in optical and infrared frequency bands, the absorption of the material is usually large at $\omega \to \omega_p$. Especially, doped semiconductors have also been proved to be able to construct ENZ media at $\omega \to \omega_p$ with relatively small loss [142, 143]. Another kind of simple ENZ media can be realized by metal waveguides, as shown in Fig. 2(b) [144-146]. When the working frequency is near the cut-off frequency of the guide mode, the effective wavelength of EM waves in the waveguide tends to infinity, thus the waveguide structure can equivalent to the ENZ media.

It should be noted that the impedance of single zero-index media (that is, ENZ or MNZ media) is seriously mismatched with the air, which is unfavorable to the high efficient transmission of EM waves. In 2011, Huang *et al.* proposed using 2D dielectric PCs to realize matched EMNZ media [147]. By adjusting the permittivity and geometric parameters of the dielectric cylinder in the 2D PC, they found that the accidental degeneracy of electric monopole and electric dipole modes can be realized, and the Dirac-like point can be formed at the center of the Brillouin zone. At this accidental degeneracy case, the 2D PC can be equivalent to an isotropic EMNZ media [147-152], which has also been successfully constructed in experiment [153] and implemented on a photonic chip [154]. As shown in Fig. 2(c), each unit consists of a silicon pillar surrounded by polymer. Due to the parallel gold film cladding, the propagating waves are completely consistent in the periodic plane, which is equivalent to the case of using silicon pillar with infinite height [154]. Interestingly, the ZIM realized by the conical dispersion can also be extended to a noncrystalline system [155].

### 2. Metal/dielectric multilayered ZIMs

Within an EMT under the condition of long-wave approximation, the periodic arrangement of artificial structures with subwavelength unit-cells can be regarded as an effective homogeneous medium, characterized by macroscopic EM parameters $\varepsilon$ and $\mu$. By designing suitable artificial structures, such as the the AMNZ media can be conveniently engineered based on the split-ring resonaotrs (SRRs) in microwave regime [156, 157]. In the visible band, metal/dielectric multilayers have been widely used to create AENZ media [158], as shown in in Fig. 2(e). According to the EMT, the effective EM parameters of metal/dielectric multilayers are [159]:

$$\varepsilon_{//} = p\varepsilon_m + (1-p)\varepsilon_d, \varepsilon_\perp = \frac{\varepsilon_m\varepsilon_d}{p\varepsilon_d + (1-p)\varepsilon_m}, \tag{1}$$

where the subscripts $\perp$ and $//$ indicate that the components are perpendicular and parallel to the $xy$ plane, respectively. $p = t_m / (t_m + t_d)$ is the filling ratio of the metal layer. $\varepsilon_i$ $(i = m, d)$ and $t_i$ $(i = m, d)$ denote the permittivity and thickness of the different layers, respectively. Especially, $\varepsilon_{//} \approx 0, \varepsilon_\perp \neq 0$ and $\varepsilon_{//} \neq 0, \varepsilon_\perp \approx 0$ correspond to two types of AENZ media with the flat IFCs along $\perp$ and $//$ directions, respectively. In order to overcome the narrow band characteristic of AENZ media constructed by metal/dielectric multilayers, an optimized stepped structure is proposed to design broadband AENZ media [160, 161]. In addition, the AENZ media have also been proposed based on the 2D PCs beyond the long-wavelength limitation [162]. This optimized EMT will facilitate the design of new metamaterials and show that the AENZ media can indeed be fabricated from a periodic 2D PC structure.

### 3. Circuit-based ZIMs

By using transmission lines (TLs), the circuit-based system can be used to construct abundant metamaterials with flexible EM parameters [163, 164]. In the circuit-based system, the relationship between the electric and magnetic fields can be easily mapped using the relationship between voltage and current. As a result, the electromagnetic response is equivalent to the circuit parameters. The structure factor of the TL is defined as $g = Z_0 / \eta_{eff}$, where $Z_0$ and $\eta_{eff}$ denote the characteristic impedance and effective wave impedance, respectively. For the general 2D TL model with lumped capacitors and resistors in Fig. 2(f), the impedance and admittance of the circuit are represented by $Z$ and $Y$, respectively. By mapping the circuit equation (telegraph equation) to Maxwell's equations, the relationship between circuit and electromagnetic parameters can be described by [163, 164]

$$\begin{aligned} \mu_{eff}\mu_0 / g &= Z / i\omega, \\ \varepsilon_{eff}\varepsilon_0 g &= Y / i\omega, \end{aligned} \tag{2}$$

where $\varepsilon_0$ and $\mu_0$ are the vacuum permittivity and permeability, respectively. The effective permittivity and permeability of the circuit system can be tuned using the lumped elements in the circuit. The above inset of Fig. 2(f) shows an effective circuit model for a circuit-based AENZ media. In this circuit model, the admittance is $Y = 2i\omega C_0$, and capacitors are loaded in the $x$ direction to realize anisotropic impedance

$$
\begin{aligned}
Z_x &= i\omega L_0, \\
Z_z &= i\omega L_0 + 1/i\omega Cd - R/d,
\end{aligned}
\tag{3}
$$

where $C_0$ and $L_0$ denote the capacitance and inductance per unit length, respectively. $d$ denotes the size of unit cell. Therefore, the effective electromagnetic parameters of the system are [163, 164]

$$
\begin{aligned}
\varepsilon &= 2C_0 g/\varepsilon_0, \quad \mu_x = \frac{L_0}{g\mu_0}, \\
\mu_z &= \frac{L_0}{g\mu_0} - \frac{1}{\omega^2 Cdg\mu_0} + i\frac{R}{\omega dg\mu_0}.
\end{aligned}
\tag{4}
$$

According to Eq. (4), we can know that the real part and imaginary part of the $\mu_z$ can be flexibly adjusted by tuning the value of the lumped capacitors $C$ and resistors $R$. Specially, when $L_0/g\mu_0 \approx 1/\omega^2 Cdg\mu_0$, the real part of $\mu_z$ is near to zero and the circuit-based AMNZ media is constructed [165, 166]. Inspired by metamaterials, researchers have recently found that resonant structures and nanoparticles can be used to design effective circuit elements in THz [167] and visible light regimes [168], thus the circuit-based ZIMs introduced here may be extended to the high frequency regimes. In addition, due to the similarity of waves, ZIMs and their novel physical properties have been successfully applied to acoustic [169-172], electronic [173] and thermal [174, 175] systems.

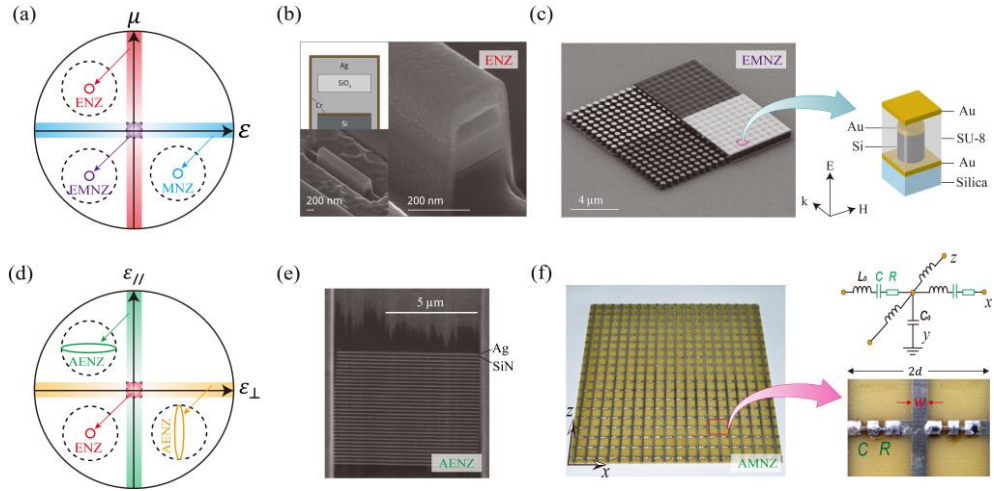

FIG. 2. **Realization of different kinds of ZIMs**. (a) EM parameter space of the isotropic ZIM. ENZ, MNZ, and EMNZ media are marked by red, blue, and purple, respectively. (b) ENZ media realized by the metallic hollow waveguide at cut-off. Reproduced with permission from Vesseur et al., Phys. Rev. Lett. 110, 013902 (2013). Copyright 2013, American Physical Society. (c) EMNZ media realized by the 2D PCs near the Dirac-like point. The corresponding unit cell is enlarged in the right inset. Reproduced with permission from Li et al., Nat. Photon. 9, 738 (2015). Copyright 2015 Springer Nature. (d) EM parameter space of the AENZ media. When the permittivity is changed to permeability, AMNZ can be obtained. (e) AENZ media realized by the subwavelength metal-dielectric multilayered structures.

## B. Physical properties

### 1. Shape independent cavity

For the conventional cavities, the resonant frequency strongly depends on the shape and size of the cavity. Generally, with the increase of the cavity size, the equivalent wavelength of the corresponding resonant mode increases, which leads to the decrease of the frequency. However, because ZIM has a very small refractive index, the wavelength of EM waves in ZIM tends to infinity. A natural question is whether ZIM can be used to design geometry-invariant resonant cavities? In 2016, Liberal *et al.*, demonstrate theoretically that ENZ media can be used to construct the deformable resonant cavities [78]. The schematics of four ZIM cavities with different shape, size or topology are shown in Fig. 3(a). Interestingly, although the geometries of these cavities are obviously different, they have the eigenmode at the same frequency. For a 2D zero-index resonant cavity in Fig. 3(b), a dielectric particle with cross-sectional area $A_i$, perimeter $L_i$, and relative permittivity $\varepsilon_i$ is embedded in the background of a ENZ media ($\varepsilon_h \approx 0$), whose cross-sectional area is $A_h$. The boundary condition of the ZIM cavity is perfectly electric conducting (PEC) wall. Under this condition, the eigenfrequencies of the ZIM cavity are determined by the characteristic equation [78]:

$$\omega = i L_i Z_S / \mu_0 A_h,\tag{5}$$

where $Z_S$ denotes the surface impedance of the particle embedded in ENZ media. From Eq. (5), we can clearly see that for a given dielectric particle, the eigenfrequencies of the ZIM cavity only depends on the area $A_h$ of the ENZ host. In other words, as long as $A_h$ is constant, no matter how the geometry of the ZIM cavity changes, it will not affect the eigenfrequency of the cavity modes. Figure 3(c) shows the simulated electric field distributions of cavity mode in four ZIM cavities with different geometry. In all these cases, the dielectric particles have similar resonance behavior. The corresponding eigenfrequency spectra of the above four ZIM cavities are shown in Fig. 3(d). It can be clearly seen that the eigenfrequency of the ZIM cavities are invariant with respect to geometrical deformations, which are marked by the red dots in Fig. 3(d). However, the quality factor $Q$ of the different ZIM cavities will changes because it depends on the special field intensity distributions in the cavities [78]. Especially, the ZIM cavities not only pave the way to design deformable resonant devices, but also provide a new platform to study the quantum optics [81, 176].

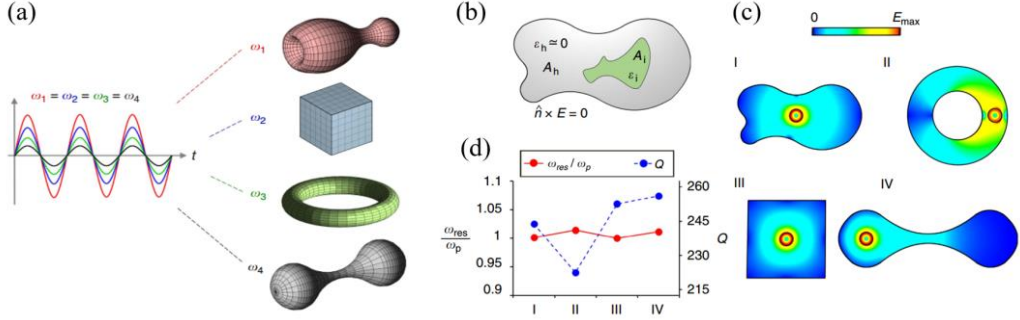

FIG. 3. **Geometry-invariant resonant cavities based on the ZIM**. (a) Schematics of four ZIM cavities with different geometry (shape, size, topology) but same resonant frequency. (b) Sketch of an irregular 2D ZIM cavity (marked by the gray region) containing a dielectric particle (marked by the green region). (c) Simulated electric field distributions of four ZIM cavities with different geometry in the vicinity of the SiC (ENZ host) plasma frequency. (d) Normalized resonance frequency and the quality factor $Q$ of four ZIM cavities with different geometry. 

### 2. Homogeneous fields

Strong fields are helpful for enhancing the interaction between light and the matter. The band gap of PCs [(AB)₅C(BA)₅] can be used to confine the EM waves and enhance the strength of the field in the defect region at the center of the PCs, which is shown in Fig. 4(a). The refractive index and thickness of different layers are ( $n_A = 1.46$ , $n_B = 2.13$ , $n_C = 1.46$ ) and ( $d_A = 100$ nm, $d_B = 100$ nm, $d_C = 200$ nm), respectively. However, the field profile is always inhomogeneous, which is mainly determined by the characteristics of standing wave field, as shown in Fig. 4(b). This leads to the limitation of some applications, such as the enhancement of nonlinear effects [74]. Based on the Maxwell equations, the fields inside the matched EMNZ media should be homogeneous. Especially, when the transverse-electric (TE) polarized waves impacting to the MNZ media ( $\mu_r \to 0$ ), the electric field is homogeneous ( $\nabla \times \vec{E} \to 0$ ) to guaranting the magnetic field is a finite value:

$$\nabla \times \vec{E} = i\omega\mu_0\mu_r\hat{H} .$$ (6)

Similarly, the magnetic field is homogeneous for the ENZ media under transverse-magnetic (TM)

polarized waves. This poses a question: can ZIM be used as special cavity to achieve uniform field enhancement?

In 2011, Jiang *et al.*, propose theoretically and demonstrate experimentally the enhancement of homogeneous fields in a zero-index cavity [74]. The schematic of a 1D PC microcavity with an effective ZIM defect is shown in Fig. 4(c), where the ZIM defect is marked by the purple region. The corresponding electric field distribution of the ZIM cavity mode is shown in Fig. 4(d). Compared with Figs. 4(b) and 4(d), we can see that the maximum values of $|E|^2$ of tranditional cavity and zero-index cavity are consistent, which are 15.4. However, the $|E|^2$ is homogenous at any place of the ZIM layer in Fig. 4(d). This enhanced uniform field boosts the average $|E|^2$ greatly, thus zero-index cavity provides a good research platform for the field enhancement without increasing the thickness of the reflecting walls. Especially, considering a nonlinear ZIM defect in Fig. 4(c), the localized fields in the nonlinear defect not only can effectively enhance the nonlinear effect, but also limit the damage of inhomogeneous field to nonlinear materials [74]. The homogenous field realized by the zero-index cavities also have been experimentally demonstrated in the microwave regime based on the circuit-based ZIMs. The photo of the sample is shown in Fig. 4(e). This zero-index cavity is made up of the electric wall (electric-single-negative media, ENG media) on the left, the magnetic wall (MNG media) on the right and the ZIM in the middle. The electric (magnetic) wall only contains ENG (MNG) units denoted by $N$ ($M$) and an effective ZIM contains a pair of ENG/MNG unit cell. The corresponding simulated electric field distribution of the cavity mode is shown in Fig. 4(f). It can be clearly seen that the electric fields are uniformly localized at the effective ZIM region and are decaying waves in the electric or magnetic wall. Moreover, the measured voltage distributions of three zero-index cavities with different size are shown in Fig. 4(g). The voltage patterns in $M_6(ZIM)_2N_6$ and $M_6(ZIM)_6N_6$ in Fig. 4(e) clearly show that the voltages in the effective ZIM region are enhanced and homogeneous [74].

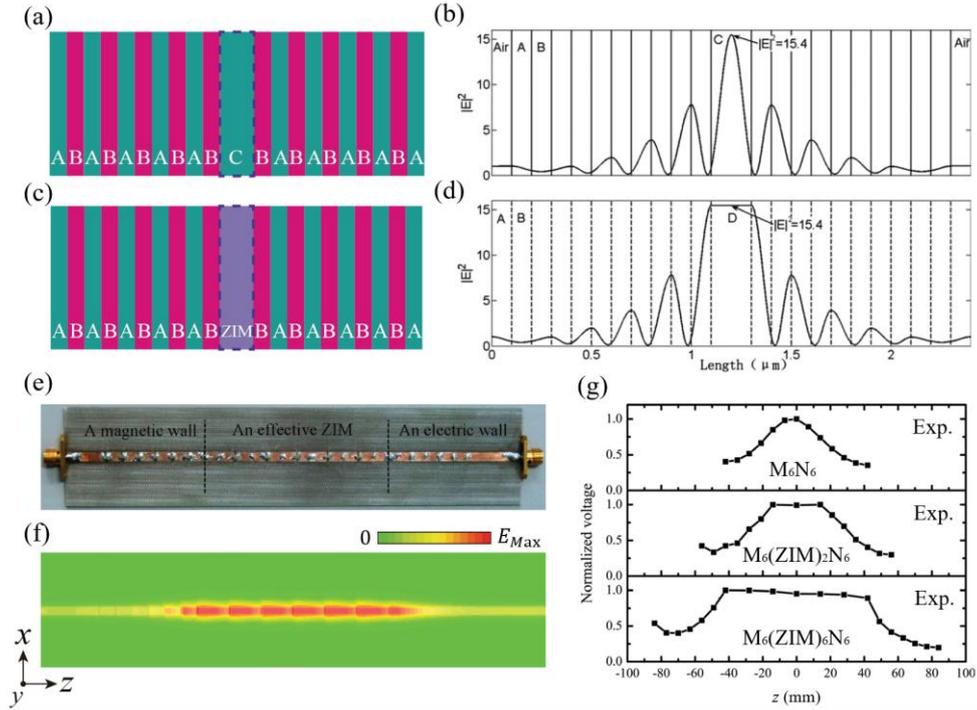

FIG. 4. **Homogeneous fields of ZIM cavity**. (a) Schematic of a 1D PC microcavity which is realized by inserting a dielectric defect (marked the layer 'C') in the center of the structure. (b) The corresponding calculated electric field distribution of the cavity mode. (c) Similar to (a), but for a 1D PC microcavity with an effective ZIM defect. (d) The corresponding calculated electric field distribution of the ZIM cavity mode. The ZIM defect is marked by the layer 'D'. (e)-(g) Homogeneous fields of the circuit-based effective ZIM cavity. (e) The sample of circuit-based ZIM cavity where the middle ZIM layer is sandwiched by an electric wall and a magnetic wall. (f) Similated electric field distribution of the 1D PC microcavity $M_6(ZIM)_6N_6$. (g) Measured voltage distributions of three effective ZIM cavities with different size: $M_6N_6$, $M_6(ZIM)_2N_6$, and $M_6(ZIM)_6N_6$. Reproduced with permission from Jiang et al., J. Appl. Phys. 109, 073113 (2011). Copyright 2011, AIP Publishing.

### 3. Inhomogeneous fields

For TM (TE) polarized waves, the electric (magnetic) field in the MNZ (ENZ) meida is a constant. Figure 5(a) shows a 2D MNZ cylindrical cavity surrounded by air [79]. When a line source with TE polarized waves is placed inside this MNZ, it can be cleary seen that the excited electric field in the MNZ media is homogeneous and the isotropic radiation appears in the air, as shown in Fig. 5(c). Interestingly, in 2015, Fu *et al.*, propose theoretically that the zero-index cavity with the air core can be used to realizing the inhomogeneous field in ZIMs [79]. This counterintuitive inhomogeneous field will appear when the higher-order modes in the zero-index cavity are excited. The schematic of a 2D MNZ shell cavity surrounded by air is shown in Fig. 5(b). Similar to Fig. 5(a), a line source with TE polarized waves is placed in the core region but off center in Fig. 5(b).

The simulated high-order cavity modes: dipole mode ($m = 0$), quadrupole mode ($m = 1$), and hexapole mode ($m = 2$) are shown in Figs. 5(d)-5(f), respectively. Compared with Fig. 5(d) to Fig. 5(f), we can clearly see the inhomogeneous field distribution of higher-order cavity modes in the zero-index cavity. The underlying physical mechanism of this property is the divergence of the magnetic field of the higher-order mode in the ZIM. According to Eq. (6), as $\mu_r H >> 0$, $\nabla \times \vec{E}$ is not zero, thus the electric feld is inhomogeneous and depends on the position, i.e., $\vec{E} = \vec{E}(r, \theta)$, where $r$ and $\theta$ denote the relative distance and angle to the center position, respectively [79]. Moreover, different from the isotropic radiation in Fig. 5(c), the controlling radiation pattern is realized based on the high-order cavity modes in ZIM, in which with numbers of outgoing direction is determined by the angular momentum $m$. The proposed ZIM cavity may be used to control (enhance or suppress) the radiation of EM waves, to control radiation pattern and to achieve isotropic or directive radiation [79].

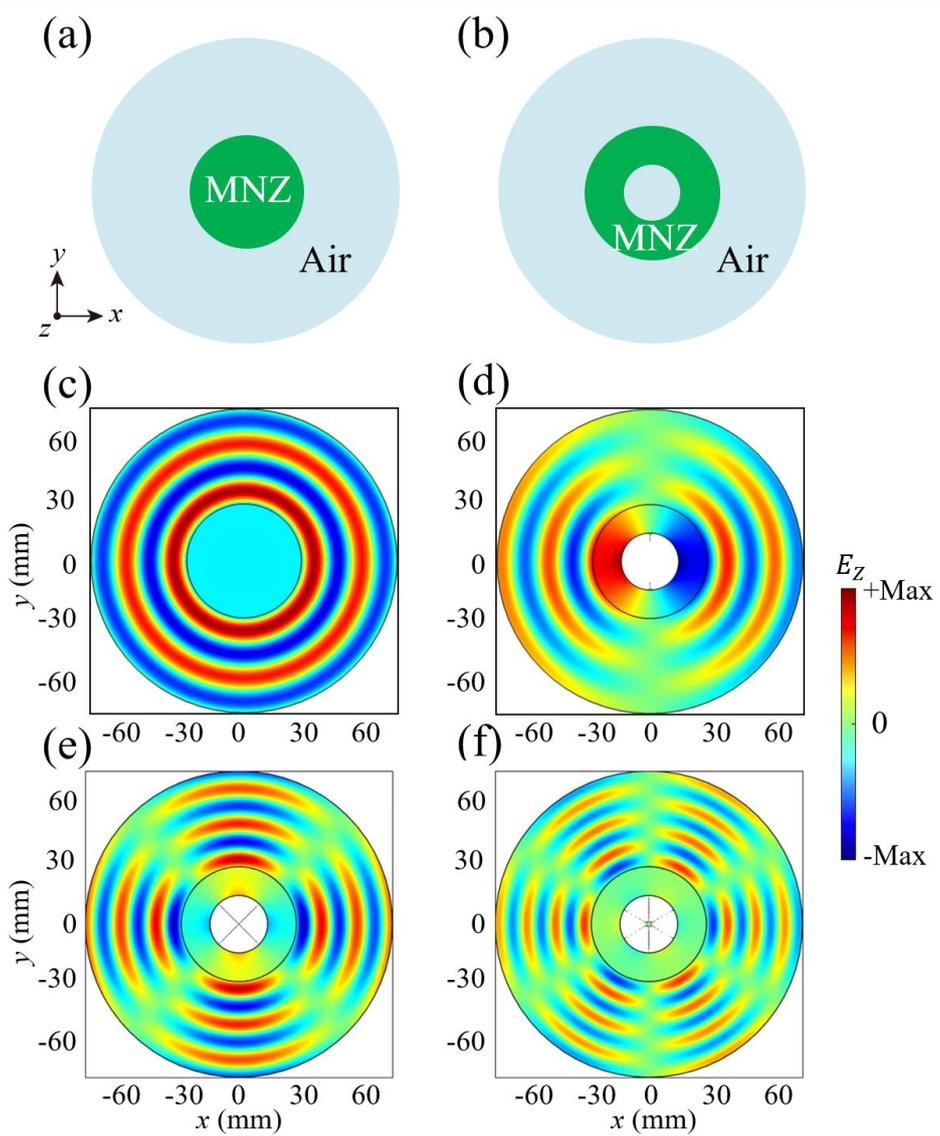

FIG. 5. **Inhomogeneous fields of ZIM cavity**. (a) Schematic of a 2D MNZ cylindrical cavity surrounded by air. (b) Schematic of a 2D MNZ shell cavity surrounded by air. (c) Simulated electric field distribution of the MNZ cylindrical cavity mode. (d)-(f) Simulated electric field distributions of the MNZ shell cavity modes for different order: (d) dipole mode ($m = 0$), (e) quadrupole mode ($m = 1$), and (f) hexapole mode ($m = 2$). 

## C. Effective zero-index (EZI) cavities

### 1. EZI cavity realized by two types of single-negative (SNG) media

In Maxwell's theory, the EM parameters of materials are characterized by permittivity $\varepsilon$ and permeability $\mu$. The appearance of metamaterials enriches the EM response of materials. People can design artificial materials with arbitrary combination of permittivity and permeability. In addition to left-handed media with $\varepsilon < 0$ and $\mu < 0$, there are also single-negative (SNG) media with negative $\varepsilon$ or $\mu$ alone. Especially, $\varepsilon < 0$, $\mu > 0$ and $\varepsilon > 0$, $\mu < 0$ correspond to ENG media and MNG media, respectively. ENG and MNG meida are painted respectively pink and blue for see in Fig. 6(a). ENG and MNG media are opaque because their refractive index is imaginary. As a result, the supported EM mode in ENG and MNG media is evanescent wave. But it is interesting that when the structure is composed of matched ENG and MNG media, there is a resonant tunneling mode, thus the combination of the two opaque materials into a new material becomes transparent [177], as shown in Fig. 6(b). The tunneling phenomenon in the heterostructure composed of ENG and MNG media can be realized under the impedance matching and phase matching conditions [177]:

$$\mathrm{Im}(Z_{ENG}) = -\mathrm{Im}(Z_{MNG}) \,, \tag{7}$$

$$\mathrm{Im}(k_{ENG})d_1 = \mathrm{Im}(k_{MNG})d_2 \,, \tag{8}$$

where $Z_{ENG}$ ($k_{ENG}$) and $Z_{MNG}$ ($k_{MNG}$) are wave impedances (wave vectors) of ENG and MNG media, respectively. $d_i$ ($i = 1$, 2) denote the thickness of the ENG or MNG layer. Considering $Z_{ENG} = \sqrt{\mu_1/\varepsilon_1}$, $Z_{MNG} = \sqrt{\mu_2/\varepsilon_2}$, $k_{ENG} = \sqrt{\mu_1\varepsilon_1}\,\omega/c$ and $k_{MNG} = \sqrt{\mu_2\varepsilon_2}\,\omega/c$, equations (7) and (8) can be can be reduced to

$$\varepsilon_1 d_1 = -\varepsilon_2 d_2 \,, \tag{9}$$

$$\mu_1 d_1 = \mu_2 d_2 \,. \tag{10}$$

Under EMT, this heterostructure can be equivalent to a matched EMNZ media with $\bar{\varepsilon} = (\varepsilon_1 d_1 + \varepsilon_2 d_2)/(d_1 + d_2) = 0$ and $\bar{\mu} = (\mu_1 d_1 + \mu_2 d_2)/(d_1 + d_2) = 0$ [177, 178]. The effective thickness of the cavity is zero because there is no phase accumulation in the cavity. The EM wave can tunnel through the pair defect satisfying Eqs. (7) and (8) without any phase delay since the pair defect is reduced to nihility [177]. Therefore, the matched ENG/MNG structure can be seen a specialy zero-index cavity, where ENG and MNG media act the mirrors.

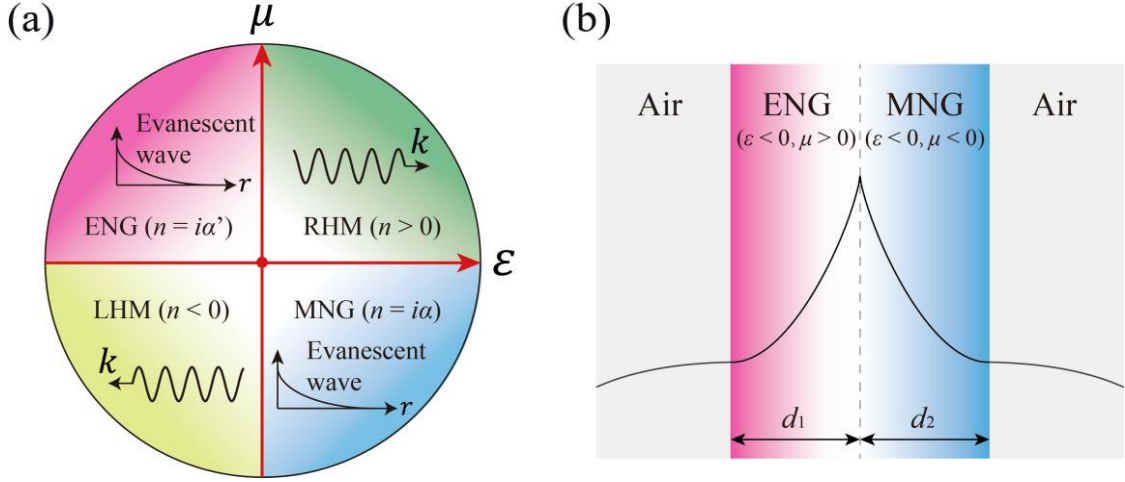

FIG. 6. **The matched pair of ENG-MNG slab**. (a) EM parameter space of the isotropic media. There are two types opaque SNG media: ENG ($\varepsilon < 0, \mu > 0$) and MNG ($\varepsilon > 0, \mu < 0$) media, which are marked by pink and blue, respectively. (b) Sketch of the field distribution of the tunneling mode in a matched pair of ENG-MNG slab.

The effective zero-index (EZI) cavity have also been widely studied in the 1D PCs [179-183]. The simplest structure is to insert a pair of ENG/MNG heterostructures directly into 1D PC [179], as shown in Fig. 7(a). The structure denotes as $(AB)_8CD(AB)_7A$, where A and B are the normal dielectric materials. C and D denote the ENG and MNG media, respectively. Especially, ENG and MNG media satisfy the matching conditions of Eqs. (7) and (8). In this case, we can find that the inserted ENG/MNG defect will not change the transmission spectrum, which is shown in Fig. 7(b). The underlying physical mechanism is come from the maintenance of wave interference. Because the ENG/MNG defect is equivalent to a transparent layer with zero effective refractive index, it has on has no effect on the interference of propagating waves in A and B layers [179]. Interestingly, although the the transmission spectrum remains invariant, the field distribution indeed changes noticeably because of the decaying wave in the pair defect. Considering the high gap-edge frequency $f_H$ in Fig. 7(b), the electric fields $\left| E(z)^2 \right|$ distributions of perfect PC: $(AB)_{15}A$ and heterostructure with ENG/MNG defect: $(AB)_8CD(AB)_7A$ are shown in Figs. 7(c) and 7(d), respectively. It can be clearly seen that the gap-edge mode of the perfect PC is a standing wave field that the electric field concentrates in the low-$\varepsilon$ regions. However, this gap-edge mode of the heterostructure with ENG/MNG defect is a propagating mode in the PC while a decaying-wave based interface mode in the pair defect, as shown in Fig. 7(d). The gap-edge mode in the zero-index cavity can be a highly localized wave instead of the usual standing wave. Especially, compared with Figs. 7(c) and 7(d), the electric field in the pair ENG/MNG defect is signicantly enhanced. Therefore, the zero-index cavity realized by matched ENG/MNG defect in a 1D PC provides a new way to control the field distribution in the structure.

In addition, Guan *et. al.*, theoretically uncover that the tunneling mode can appear from any other combined structure composed of MNG and ENG slabs so long as the general zero average conditions ($\bar{\varepsilon} = 0$, $\bar{\mu} = 0$) are satisfied [180]. Figure 7(e) shows a heterostructure constituted by two different 1D PCs: (CD)$_m$ and (C'D')$_n$ with SNG media, where C (C') and D (D') represent the ENG and MNG media, respectively. The general zero average conditions of the tuneling mode in the heterostructure can be written as

$$\bar{\varepsilon} = \frac{m(\varepsilon_C d_C + \varepsilon_D d_D) + n(\varepsilon_{C'} d_{C'} + \varepsilon_{D'} d_{D'})}{m(d_C + d_D) + n(d_{C'} + d_{D'})} = 0 , \tag{11}$$

$$\bar{\mu} = \frac{m(\mu_C d_C + \mu_D d_D) + n(\mu_{C'} d_{C'} + \mu_{D'} d_{D'})}{m(d_C + d_D) + n(d_{C'} + d_{D'})} = 0 . \tag{12}$$

It can be clearly seen that there is a band gap exist in the individual left PC: (CD)$_m$ and right PC: (C'D')$_n$ in Fig. 7(f). However, when the zero average conditions satisfied for the heterostructure: (CD)$_m$(C'D')$_n$, the tunneling mode with unit transmittance appears inside the forbidden gap. The corresponding electric fields $\left| E(z)^2 \right|$ distributions of the tunneling mode is shown in Fig. 7(g). In this case, the electric field is mainly localized at the interface of the two PCs. Especially, this unusual tunneling mode realized by the EZI cavity is independent of incident angles and polarizations and have zero phase delay, which can be utilized to design zero-phaseshift omnidirectional filters [180].

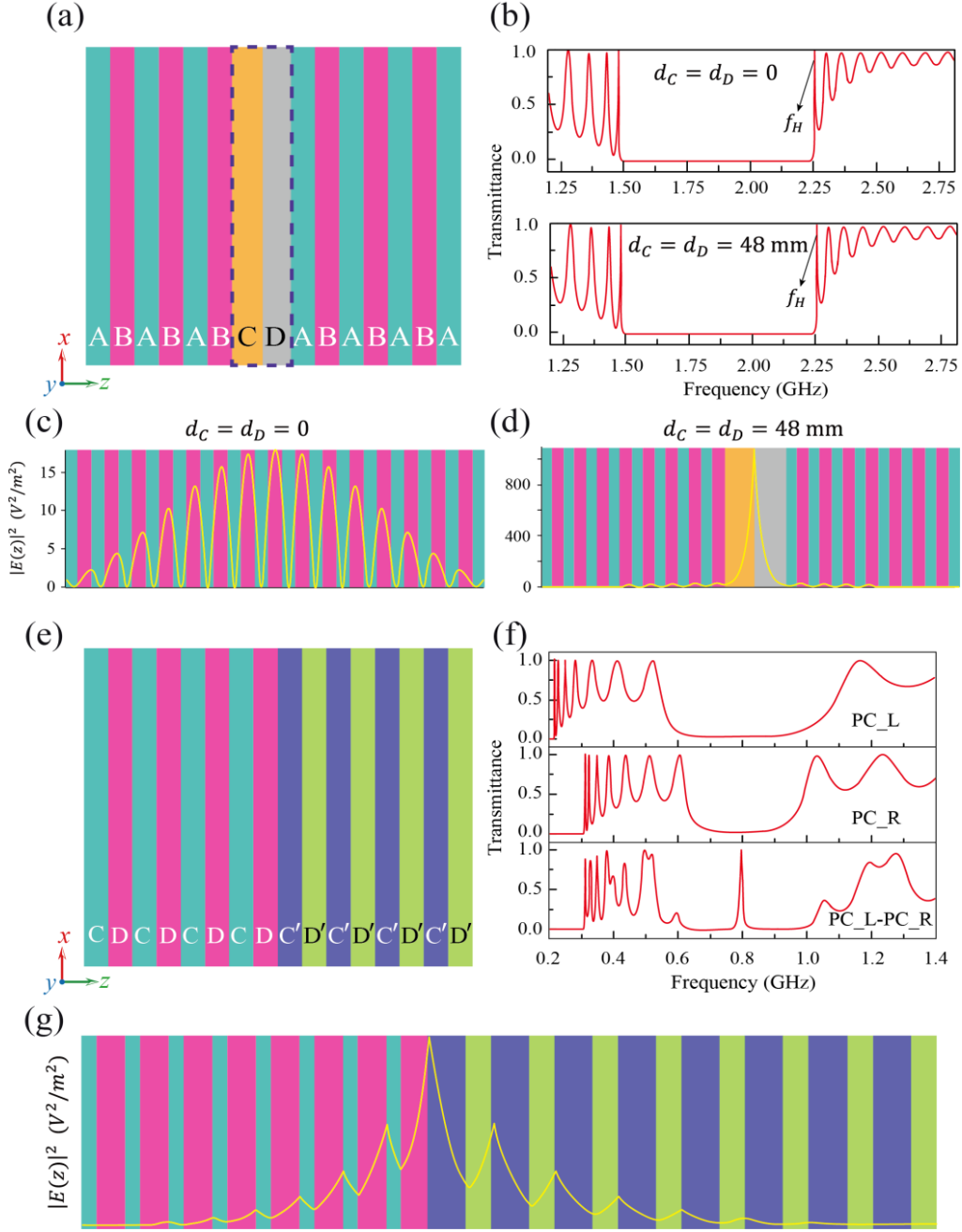

FIG. 7. **Realization of EZI cavity based on two types of SNG media**. (a) Schematic of a 1D PC with ENG and MNG defects, which are marked by the layers C and D, respectively. (b) Transmittance of the perfect PC: $(AB)_{15}A$ and heterostructure: $(AB)_8CD(AB)_7A$. (c) (d) Distribution of $|E|^2$ in the heterostructure $(AB)_{15}A$ and $(AB)_8CD(AB)_7A$ at frequency $f_H$ as indicated in (b). <span style="color:blue">Reproduced with permission from Jiang et al., Phys. Rev. E 73, 046601 (2006). Copyright 2006, American Physical Society.</span> (e) Schematic of a heterostructure constituted by two different 1D PCs $[(CD)_m$ and $(C'D')_n]$, where C (C') and D (D') denote the ENG and MNG media, respectively. (f) Transmittances of the structures $(CD)_8$, $(C'D')_8$, and $(CD)_8(C'D')_8$. (g) Electric field distribution corresponding to the tunneling mode at 0.7958 GHz. <span style="color:blue">Reproduced with permission from Guan et al., Appl. Phys. Lett. 88, 211112 (2006). Copyright 2006, AIP Publishing.</span>

## 2. Enhanced magneto-optic effect and nonlinear effect

In general, the large magneto-optical (MO) activity and opotical nonlinearity of materials are highly desirable in many applications, such as the isolators, optical switches, etc. In this section, we review that the EZI cavity can be used to significant enhance the MO effect and nonlinear effect. Compared with the single-layer MO medium, the optical isolator realized by MO PC has the advantages of high transmission, strong Faraday rotation effect and small volume. Transparent yttrium iron garnet (YIG) is one of the most studied MO media. Although the MO activity of MO metal is much larger than that of YIG, the MO metal is opaque, thus the MO activity of MO metal is nearly inaccessible. Similarly, the the third-order nonlinear susceptibility of noble metals is large, but it is difficult to be used. Interestingly, the opaque metal with MO activity or opotical nonlinearity can become transparent when the heterostructure (i.e., EZI cavity) is formed by matching suitable materials. In addition, the tunneling mode in EZI cavity can realize the strong localization of the field, thus enhancing the interaction between light and matter.Considering a heterostructure consisting of an all-dielectric PC: (AB)$_6$ and a MO metal layer (M), the EZI cavity with tunneling mode is constructed [75]. The corresponding schematic of heterostructure (AB)$_6$MP is shown in Fig. 8(a). B is the SiO$_2$ layer with $\varepsilon = 2.1$. P is the protection film SiO2. A and M are the MO media, which correspond respectively to Bi : YIG and Co$_6$Ag$_{94}$ media. Under an applied magnetic field is in the $z$ direction, the permittivity tensor of layer A and M is:

$$\bar{\varepsilon} = \begin{pmatrix} \varepsilon_{xx} & i\alpha & 0 \\ -i\alpha & \varepsilon_{yy} & 0 \\ 0 & 0 & \varepsilon_{zz} \end{pmatrix}, \tag{13}$$

where $\alpha$ is the off-diagonal element responsible for the strength of MO activity of the medium. The transmission (solid line) and reflection (dashed line) spectrum of the heterostructure (AB)$_6$MP is shown in Fig. 8(b). There is a dip of zero reflection at $\lambda = 631$ nm, that is the the tunnelling mode satisfied the matched consitions [75]:

$$\mathrm{Im}(Z_{PC}) = -\mathrm{Im}(Z_M), \tag{14}$$

$$6\mathrm{Im}(k_{PC})(d_A + d_B) = \mathrm{Im}(k_M)d_M. \tag{15}$$

As mentioned above, the impedance matching and phase matching conditions are equivalent to the zero average conditions. Therefore, the heterostructure composed of PC and metal layer also belongs to the EZI cavity. Especially, the correspoinding spectrum of the Faraday rotation angles $\theta_f$ is shown in Fig. 8(c). It is seen that the maximal $\theta_f$ emerges at the wavelength of the tunnelling mode $\lambda = 631$ nm. Considering the tunnelling mode, the electric fields $\left| E(z)^2 \right|$ distribution of the heterostructure $(AB)_6MP$ is shown in Fig. 8(d). It can be clearly seen that the electric field of the tunnelling mode is strongly localized at the interface between MO PC and the MO metal layer. It is the the slow-wave effect of the localized electric field in the MO metal that the Faraday rotation effect is enhanced. For comparison, figures 8(e)-8(h) show the results of a tranditional metallo-dielectric magnetophotonic crystal. The metallo-dielectric magnetophotonic crystal $(A'M')_5$ is composed of dielectric and MO metal, where A' and M' denote respectively $TiO_2$ and MO metal $Co_6Ag_{94}$, as shown in Fig. 8(e). Especially, the total thickness of the M' layer in Fig. 8(e) is the same as that of the M layer in the heterostructure $(AB)_6MP$ in Fig. 8(a). By choosing appropriate parameters, the short band-edge mode $\lambda = 627$ nm can be close to the tunneling mode in Fig. 8(b), and the system has high transmittance $T$ and Faraday rotation angle $\theta_f$. Similar to Figs. 8(b) and 8(c), figures 8(f) and 8(g) give the $T$ and $\theta_f$ spectrum of the tranditional magnetophotonic crystal $(A'M')_5$, respectively. The corresponding of the electric fields $\left| E(z)^2 \right|$ distribution is shown in Fig. 8(h). The nodes of the electric field locate at each thin silver layer and the strength of the electric field in all the silver layers in Fig. 8(h) is less than that of the heterostructure in Fig. 8(d). As a result, the enhancement of MO effect of the heterostructure $(AB)6MP$ is larger than that in the metallo-dielectric photonic crystal $(TiO2/Co6Ag94)5$ due to the higher electric field configuration in the MO metal. The large MO effect realized by the EZI cavity may be used to design compact MO devices [75].

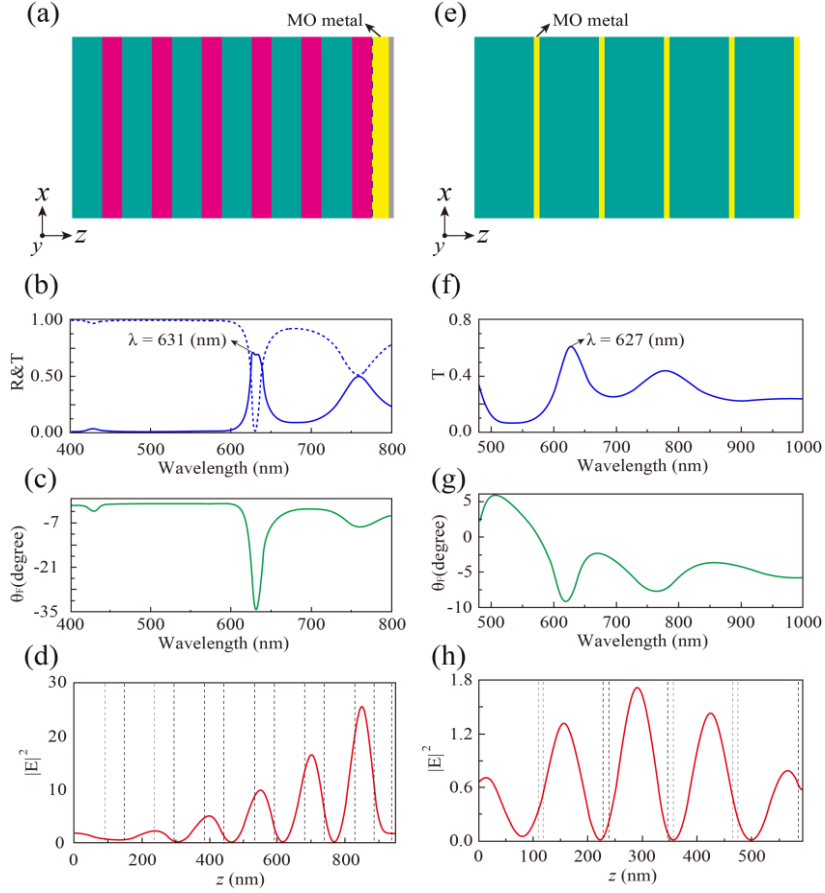

FIG. 8. **Enhanced Faraday rotation effect based on the EZI cavity**. (a) Schematic of a 1D heterostructure $(AB)_6MP$ composed of an all-dielectric photonic crystal $(AB)_6$, a MO metal M (painted by the yellow) and a protection film P (painted by the gray) in air. (b) Reflectance $R$ (dashed line) and transmittance $T$ (solid line) of the heterostructure $(AB)_6MP$. The working frequency is selected to be 631 nm, which is marked by the black arrow. (c) Faraday rotation angle $\theta_F$ of the heterostructure $(AB)_6MP$. (d) Distribution of $|E|^2$ in the structure at $\lambda = 631$ nm as indicated in (b). (e)-(h) Similar to (a)-(d), but for the 1D metallo-dielectric magnetophotonic crystal: $(A'M')_5$. The working frequency is selected to be 627 nm, which is marked by the black arrow in (f). Reproduced with permission from Dong et al., J. Phys. D: Appl. Phys. 44, 145402 (2011). Copyright 2011, Institute of Physics.

In addition to the MO effect, the EZI cavity can also enhance the nonlinear effect significantly. For a heterostructure composed of an all-dielectric PC and a metallic film with nonlinearity, the tunneling mode can produce large nonlinear effect because of the strong field localization. The schematic of a 1D heterostructure $(AB)_7M$ composed of an all-dielectric PC: $(AB)_6$ and a metal: Ag layer is shown in Fig. 9(a). A and B denote $SiO_2$ and $TiO_2$ with permittivity $\varepsilon_A = 2.1$ and nD=5.4, respectively. The permittivtiy of silver with nonlinearity can be written as [184]:

$$\varepsilon_{Ag}^{NL} = \varepsilon_{Ag}^{L} + \varepsilon_0\chi_3|E|^2, \tag{16}$$

where $\varepsilon_{Ag}^{L} = 1 - \omega_p^2/(\omega^2 + i\gamma\omega)$ is the linaer permittivity and $\chi = 2.4\times10^{-9}$ denotes the nonlinear susceptibilities of silver. Figure 9(b) shows the transmission spectrum of the heterostructure $(AB)_7M$ without considering the nonlinear susceptibilities of silver (i.e., $\chi = 0$). It can be clearly seen a tunneling mode with frequency $f_0 = 525$ THz appears, which is marked by the dashed line. The corresponding electric and magnetic fields of the tunneling mode are shown in Fig. 9(c). Especially, the EM fields are localized at the interface between the $SiO_2/TiO_2$ PC and the silver layer. Considering the nonlinear susceptibility of silver, the electric field in silver and the frequency of the tunneling mode varies with the electric field intensity $|E|_{in}^2$ of incident wave. Because the strong field localization induced large nonlinear effect in the EZI cavity, the bistability can be observed near the frequency of tunneling mode. Figure 9(d) shows the nonlinear properties of the heterostructure at three different frequencies: 521.5 THz (pink line), 520.6 THz (orange line), and 519.5 THz (green line). For the case that the frequency is 521.5 THz, the bistability can be clearly observed [184]. Especially, $|E|_{in1}^2$ and $|E|_{in2}^2$ represent the intensities of switching-up and switching-down thresholds for bistability, respectively. Moreover, the variance of intensities of thresholds with the frequencies of the incident wave is shown in Fig. 9(e). It can be clearly seen that 520.6 THz corresponds to critical frequency of the incident wave and there is no bistability above 520.6 THz. For comparison, figures 9(f)-9(j) show the results of a tranditional metallo-dielectric PC: $(SiO_2Ag)_7$. Especially, the total thickness of the Ag layer in Fig. 9(f) is the same as that of the Ag layer in the heterostructure in Fig. 9(a). Similar to Figs. 9(b), figures 9(g) give the $T$ spectrum of the tranditional metallo-dielectric PC. The frequency of the band-edge mode in Fig. 9(g) is designed equal to the frequency of the tunneling mode in Fig. 9(b). And the corresponding $|E|^2$ distribution of the band-edge mode is shown in Fig. 9(h). It is seen that the nodes of the electric field locate at each thin silver layer, thus the tranditional metallo-dielectric PC can enhance transmittance, but the electric field in silver is still weak, which limits the enhancement of the nonlinear effect [184]. The nonlinear response and the threshold strength of 1D tranditional metallo-dielectric PC is shown in Fig. 4(i) and 4(j), respectively. Compared with Fig. 9(e) and 9(j), the critical intensity of the threshold in the EZI cavity is reduced by nearly 2 orders of magnitude than the tranditional metallo-dielectric PC, which may be used in in many applications, such as bistable switching, light-emitting diodes, etc [184].

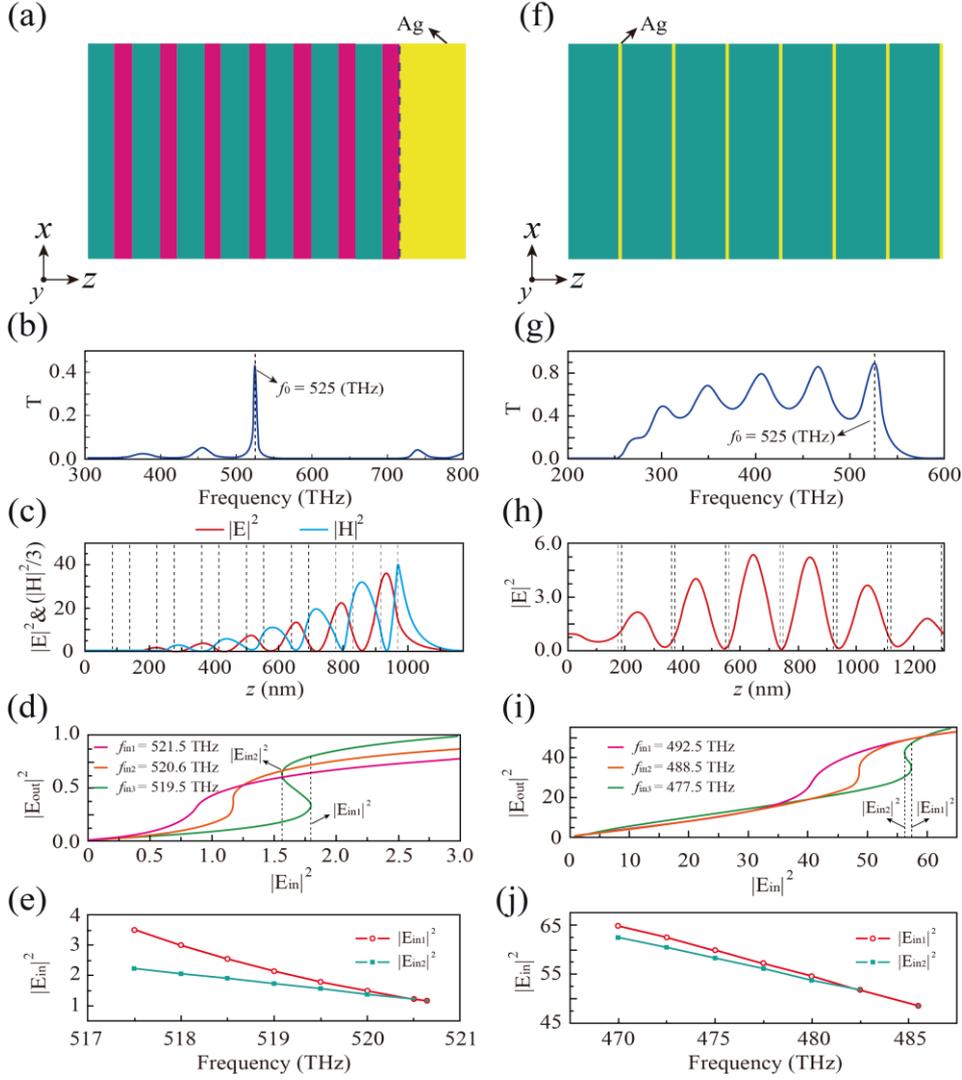

FIG. 9. **Enhanced optical nonlinearity based on the EZI cavity**. (a) Schematic of a 1D heterostructure $(AB)_7M$ composed of an all-dielectric PC: $(AB)_6$ and a metal: Ag layer with nonlinear susceptibilities. (b) Transmittance of a heterostructure $(AB)_7M$ without considering nonlinear susceptibilities of Ag. (c) Distributions of $|E|^2$ and $|H|^2$ in the heterostructure at the frequency $f = 525$ THz as indicated in (b). (d) Output versus input intensity at three different frequencies: 521.5 THz, 520.6 THz, and 519.5 THz. (e) Intensities of thresholds for bistability versus frequencies of the incident wave. (f)-(j) Similar to (a)-(e), but for the 1D metallo-dielectric PC. The working frequency of (h) corresponds 525 THz, which is marked by the black arrow in (g). Reproduced with permission from Du et al., Opt. Lett. 34, 578 (2009). Copyright 2009, OSA Publishing.

### 3. EZI cavity with topological characteristics

Recently, the topological photonics has attracted people's great attention due to their great advantages in fundamental topological research and practical applications [185-187]. Different from the tight-binding model, 1D PCs with multiple scattering mechanism are also the important topological structures [188-192]. In which, the topological properties can be easily affected by the geometrical settings. Especially, the EM response of materials depends on the permittivity and permeability. When one of them is negative, the material corresponds to the SNG media and they can be seen the light mirrors. By mapping the 1D Maxwell equation to 1D Dirac equation, the topological order of material can be determined by the effective mass $m$ associated with the effective permittivity and permeability [190]:

$$[-i\sigma_x\partial_x + m(x)\sigma_z + V(x)]\varphi = E\varphi, \tag{17}$$

where $\sigma_x$ and $\sigma_z$ are the Pauli matrices. $m(x) = \omega[\varepsilon_r(x) - \mu_r(x)]/2c$ and $V(x) = \omega[\varepsilon_r(x) + \mu_r(x) - \langle \varepsilon_r(x) + \mu_r(x)\rangle]/2c$ are the effective mass and effective potential, respectively. The topological properties of bands or the band gaps can be directly distinguished by the effective EM parameters [193, 194]. For ENG media, $\varepsilon$ is negative, $\mu$ is positive and the effective mass $m$ is negative. However, for MNG media, $\varepsilon$ is positive, $\mu$ is negative and the effective mass $m$ is positive. Based on this method, the topological edge states in the heterostructure composed of two PCs with different topological orders have been proposed theoretically in visible light band and experimentally verified in microwave band [188-190]. So the topological order of electric and magnetic mirrors are different. The circuit system based on the TLs provides a good platform to study the topological structure and the related properties. Figure 10 (a) shows the photograph of a paired structure, which composed of an circuit-based ENG media and a PC [188]. Once the right PC can be effective to MNG media, the EZI cavity with tunneling mode can be realized. In addition, considering two different PCs, the left PC produces the effect of an ENG media and the right that of a MNG media in Fig. 10(b). In Fig. 10(c), the transmission spectra of the right PC, the left PC, and the paired structure are shown by the blue dotted lines, the red dashed lines, and the solid black lines, respectively. For the paired structure, a tunneling mode is at 2.91 GHz in simulation and 3.07 GHz in the experimental measurement, as indicated by the green dotted lines in Fig. 10(c) [188]. Especially, by tailoring the permittivity and permeability of metamaterials, band inversion of the Dirac equation was demonstrated theoretically and experimentally [190]. It has been found that the band inversion accompanies a change of chirality of electromagnetic wave in metamaterials. Three samples: left PC, right PC and the paired structure are constructed based on the circuit-based system. The density of states (DOS) of the paired structure is simulated and measure in the left and right panels of Fig. 10(e), respectively. The topological edge state (i.e., the tunneling mode in the EZI cavity) is identified by the additional narrow peak appears at $\omega = 11.05$ GHz within the gap region in Fig. 10(e). Figure 10(f) shows the full-wave simulation of field spatial distribution of the topological edge state. It is clearly seen that the topological edge state is strongly localized at the interface of two PCs with different topological orders. In addition, the measured voltage distribution is meet well with the simulated field distribution, as shown in Fig. 10(g). The EZI cavities with topological characteristics not only provide a proof-of-principle example that EM wave in the metamaterials can be used to simulate the topological order in condensed matter systems, but also are helpful for research into surface modes in PCs and related applications [188-190].

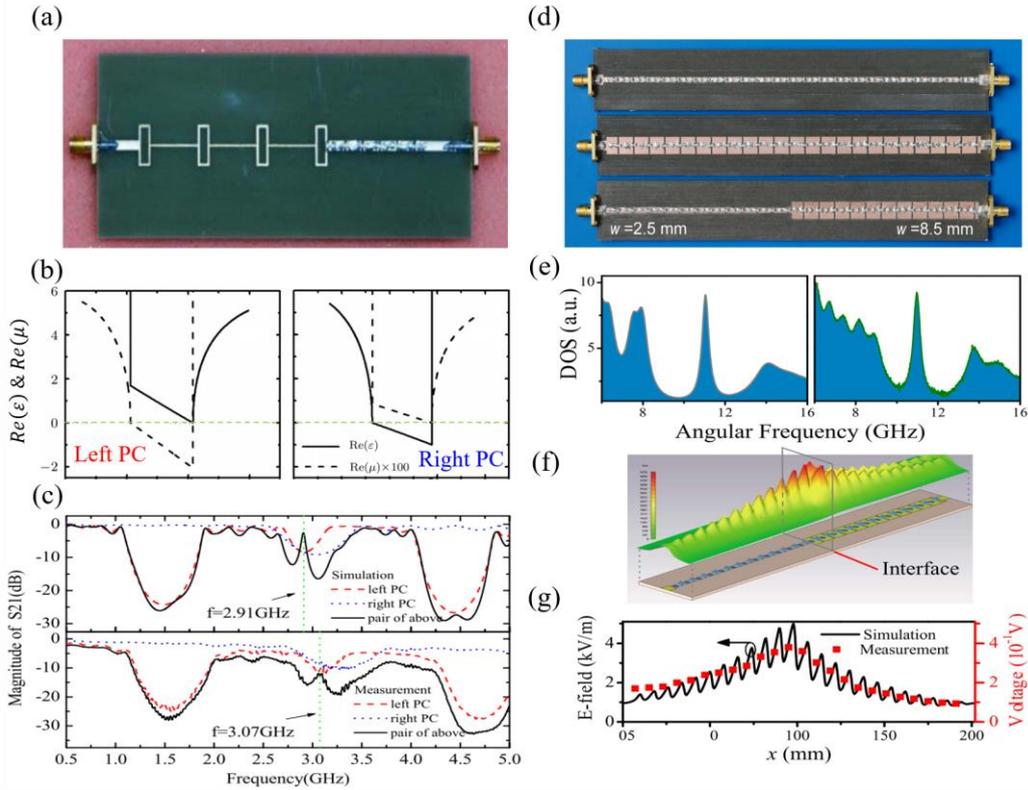

FIG. 10. **Effective parameters and topological order (effective mass) of the band gaps**. (a) Photograph of a paired structure based on the TLs. (b) The efiective parameters of the band gaps of two PCs. The band gaps of left and right PCs correspond to the ENG ($\varepsilon < 0$, $\mu > 0$) medium and the MNG ($\varepsilon > 0$, $\mu < 0$) medium, respectively. (c) Simulated and measured transmissions of left PC (the red dashed line), right PC (the blue dotted line), and their paired structure (the black solid line). The edge states of the EZI cavity are marked by the black arrows. Reproduced with permission from Guo et al., Phys. Rev. E 78, 026607 (2008). Copyright 2008, American Physical Society. (d)-(g) Edge mode at the interface between two photonic mirrors with $m > 0$ and $m < 0$. (d) Three fabricated samples demonstrating the domain wall and edge state. (e) Calculated (left panel) and measured (right panel) DOS spectra, exhibiting an edge state within the band gap. (f) Full-wave simulation of field distribution of the edge state. (g) Measured voltage distribution of the edge state in the heterostructure composed of two photonic mirrors with different topological order (effective mass). Reproduced with permission from Tan et al., Sci. Rep. 4, 3842 (2014). Copyright 2014 Springer Nature.

The photonic topological edge states based on the 1D PCs have also been demonstrated at high frequency regimes. For the a 1D PC with symmetric unit cell in Fig. 11(a), the topological property of the bands can be determined by the Zak phase [191]. The unit cell is marked by the dashed yellow lines. The light is incident from the left side of the structure. Figure 11(b) shows the topological edge state realized by a heterostructure composed of two topological distinguished PCs. The left PC1 and right PC2 are denote respectively by $(A_1B_1A_1)_{10}$ and $(A_2B_2A_2)_{10}$, where $\varepsilon_{A1} = 3.8$, $d_{A1} = 0.21d$, $\varepsilon_{B1} = 1$, $d_{B1} = 0.58d$, $\varepsilon_{A2} = 4.2$, $d_{A2} = 0.19d$, $\varepsilon_{B2} = 1$, $d_{B2} = 0.62d$. $d = 2d_A + d_B$ is the thickness of the unit cell. The band structure of the 1D PC can be obtained by:

$$\cos(qd) = \cos k_A d_A \cos k_B d_B - \frac{1}{2}(\frac{z_A}{z_B} + \frac{z_B}{z_A})\sin k_A d_A \sin k_B d_B ,  \qquad (18)$$

where $q$ represents the Bloch wavevector. Based on the above parameters, the corresponding band structures of PC1 and PC2 are shown in the middle and right panels of Fig. 11(b). Especially, the Zak phase (0 or $\pi$) of bands is marked near the bands. The topological property of the bandgap is painted different colors for see. It can be clearly seen that the 7th band gap of two PCs is topological distinguished. As a result, a topological edge state will occur in the heterostructure PC1-PC2 at the 7th band gap, which can be observed at the calculated transmission spectrum in the left panel of the Fig. 11(b).

Recently, the short-wavelength optical science is undergoing great development. In fact, the refractive indices of all materials are close to 1 in X-ray band, so a single layer cannot be used as a photonic insulator. The X-ray insulator is realized by using the band-gap of multilayer structures. However, in which the size of unit cell is close to the atomic scale. So the structural fluctuations are unavoidable. A question naturally arises: can the topological properties be applied in the X-ray band to enable new devices? Recently, the topological edge state based on the 1D PCs have also been extended to the the X-ray band [192]. To obtain a gap with strong reflectance, the X-ray is at a grazing incident angle. Considering a PC with symmetric unit cell, the characteristic matrix can be considered as a single layered material because two diagonal matrix elements are equal. To simplify the fabrication, two topological distinguished PCs are constructed by $(CWC)_{10}$ and $(WCW)_{10}$, respectively. The thickness of each layer is same $d_{A1} = d_{B1} = d_{A2} = d_{B2} = 1.5$ nm. The effective EM parameters of PC1 and PC2 are shown in Fig. 11(c) and 11(d), respectively. For the in the grazing incidence angular range of $\alpha \in (1.05^\circ, 1.1^\circ)$, PC1 and PC2 correspond to the ENG and MNG media, respectively. Based on these effective parameters, the topological order of the band gap can be determined base on the effective mass $m$. In Fig. 11(e), the reflection phases of PC1 and PC2 belong to the ranges of $(-180°, 0)$ and $(0, 180°)$, which further confirms that the bandgaps of PC1 and PC2 are ENG and MNG gaps, respectively. For the edge state in the EZI cavity, the field maximum is at the interface of the two kinds of PCs. Moreover, from the interface to the left or right end of the structure, the envelope of the field exponentially decays, as shown in Fig. 11(f).

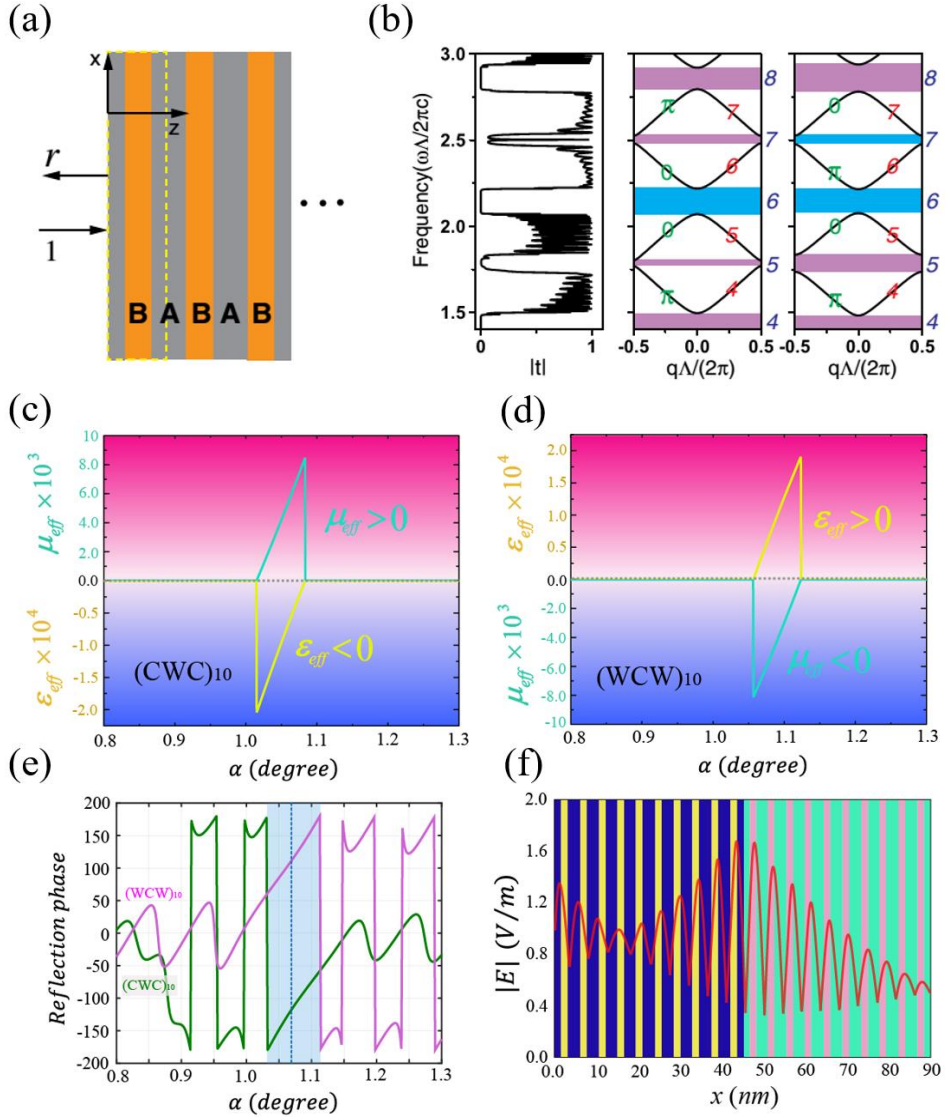

FIG. 11. **Reflection phase of the band gaps and the topological order (Zak phase) of the bands of PCs**. (a) Schematic of a 1D PC with symmetric unit cell, which is marked by the dashed yellow lines. (b) Transmission of the 1D heterostructure composed of PC1 and PC2, and the band structures (solid black curve) of single PC1 and PC2. The Zak phase of each individual band is labeled in green. Reproduced with permission from Xiao et al., Phys. Rev. X 4, 021017 (2014). Copyright 2014 Author(s), licensed under a Creative Commons Attribution 3.0 License. The effective parameters of PC1 (c) and PC2 (d) in X ray band, which correspond to the ENG and MNG metamaterials, respectively. The unit cell of PC1 and PC2 are CWC and WCW, respectively. (e) The reflection phases of PC1 (green line) and PC2 (purple line). The band gap is marked by the blue shaded region. (f) Electric field distribution of the heterostructure composed of PC1 and PC2, where PC1 is on the left-side and PC2 is on the right-side of the interface. Reproduced with permission from Huang et al., Laser & Photon. Rev. 13, 1800339 (2019). Copyright 2019 Wiley-VCH.

The transmission electron microscope (TEM) image of the fabricated PC1–PC2 sample is shown in Fig. 12(a) [192]. The interface region is measured in high-resolution images, as shown in Fig. 12(b). The bright layers are carbon and the dark layers are tungsten. For the individual PC1 and PC2, they will exhibit a Bragg gap. However, once the topologically distinct PC1 and PC2 are combined together, there is a reflection dip in the X-ray bandgap. Which is the topological edge state. Remarkably, this topologically protected edge state is immune to the thickness disorder as long as the zero-average-effective-mass ($\bar{m} = 0$) condition is satisfied [190, 192]:

$$\bar{m} = \int_{-L_1}^{0} m_1(x)dx + \int_{0}^{L_2} m_2(x)dx = 0 , \qquad (19)$$

where $m_1(x)$ and $m_2(x)$ are the effective masses of the unit cell of PC1 and PC2, respectively. Length $L_1$ and $L_2$ denote the total thicknesses of PC1 and PC2, respectively. The interface position of the two PCs is defined as 0. By adding certain kinds of disorders into PC2, the robustness of X-ray edge state in the hetero-structure is demonstrated in Figs. 12(c)-12(f). It can be seen that when the topological protection condition is still satisfied, the edge state is almost the same as that without disorder. On the contrary, if this condition is not satisfied, the edge state will be greatly affected. Therefore, based on two kinds of 1D PCs with different topological properties, the topological edge state in the X-ray band is demonstrated. Imporantly, this edge state is demonstrated to be robust against thickness disorders as long as the zero-average-effective-mass condition is satisfied. The related results extend the concept of topology to the X-ray band and may provide insightful guidance to the design of novel X-ray devices with topological protections [192]. In addition, the EZI cavity with topological properties also has been extended to the acoustic system [195]. In general, the combination of topology and ZIM will produce more interesting physical properties [196, 197], which is worthy of further study in the future.

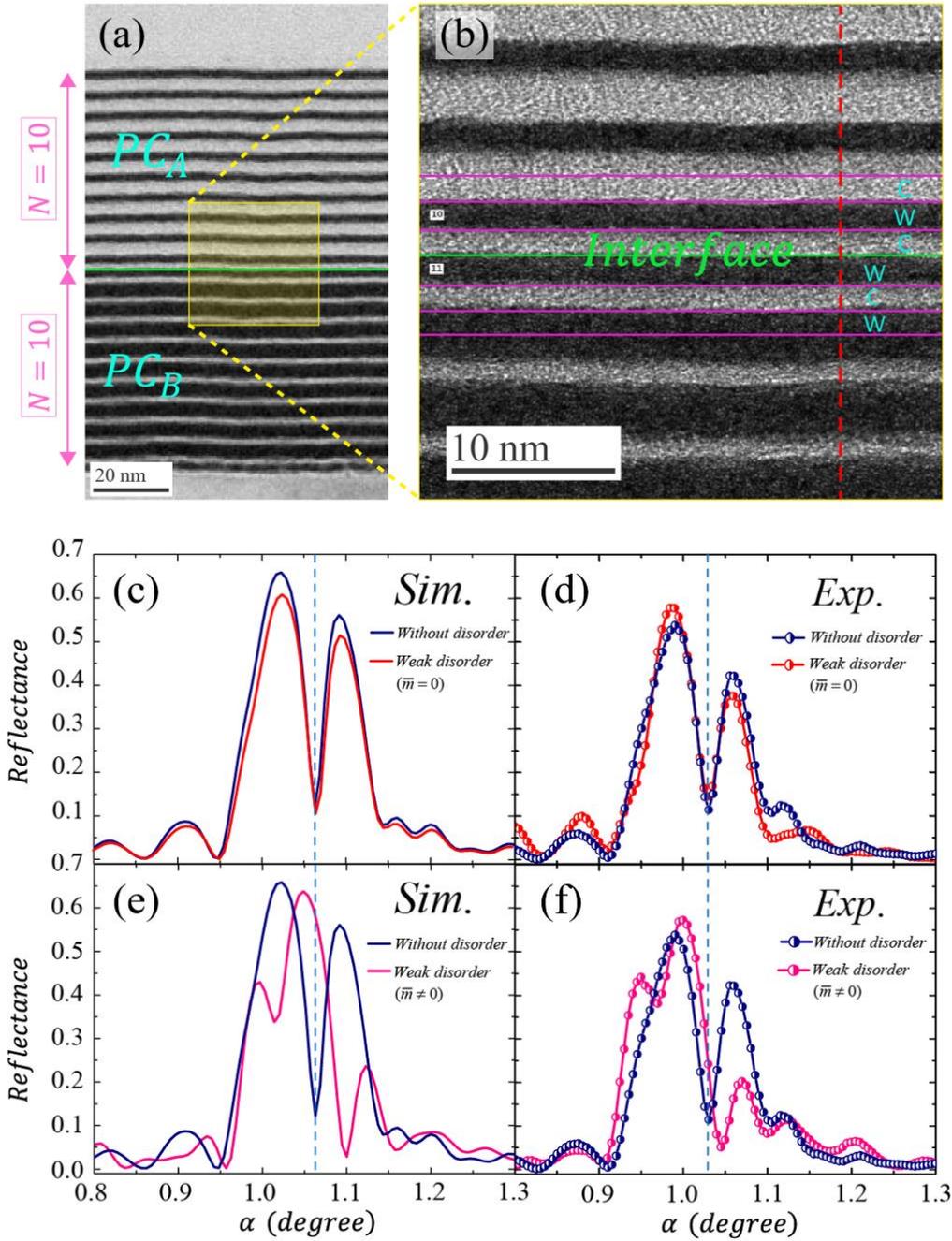

FIG. 12. **Demonstration of the topological edge state associated with the EZI cavity.** A TEM image of the cross-section of the fabricated PC1–PC2 structure. An enlarged image of the interface area is shown in (b), in which the interface of PC1 and PC2 is marked by the green line. Simulated (c) and measured (d) reflectance of the heterostructure with a thickness disorder in which $\bar{m} = 0$. (e) (f) Similar to (c) (d), but for the heterostructure with a thickness disorder in which $\bar{m} \neq 0$. Reproduced with permission from Huang et al., Laser & Photon. Rev. 13, 1800339 (2019). Copyright 2019 Wiley-VCH.

### D. Applications

#### 1. Switching

Active control of EM wave transmission and radiation greatly enriches the flexibility of light control. In 2018, Neira *et al*., theoretically proposed that all-optical switching can be realized by zero-index cavity based on the strong nonlinearity [83]. The switchable ENZ cavity is constructed by two ENZ mirrors, which are implemented in the silicon waveguide, as shown in Fig. 13(a). According to the EMT, the effective anisotropic ENZ layer is realized by using gold nanowire array [83, 198]:

$$\varepsilon_z = N\varepsilon_{Au} + (1-N)\varepsilon_{Si} \, , \qquad (20)$$

where $\varepsilon_{Au}$ and $\varepsilon_{Si}$ are the permittivity of Si and Au, respectively. The concentration of gold nanowire is defined by the interdistance $p$ and nanorod diameter $2a$ as $N = \pi(a/p)^2$. Especially when the cavity mode frequency is tuned to that of the effective ENZ media, the high and low transmission of the structure is obviously affected by the ENZ condition [83]. For the high-transmission state at ENZ condition, the cavity mode is built up between two layers of nanorods. However, for the low-transmission state deviation from ENZ condition, the mode is mainly reflected. In addition, as mentioned in the Sec. C, the nonlinear response of the materials can be improved significantly by ZIMs. Therefore, the optical modulator based ons the zero-index cavity provides a double advantage of high mode transmission and strong nonlinearity enhancement in the few-nanorod-based design. In the infrared regime, the permittivity of Au can be written as [83]:

$$\varepsilon_{Au} = \varepsilon_\infty - \frac{\omega_p^2(T_L)}{\omega[\omega + i\gamma_{Intra}(\omega, T_L, T_e)]} \, , \qquad (21)$$

where $\gamma_{Intra}(\omega, T_L, T_e)$ is the electron scattering rate, including both electron–electron and electron–phonon scattering as a function of the temperature of electrons $T_e$ and lattice $T_L$, respectively. The modulation is determined by the control pulse ($\sim$100 fs duration), which changes the electron temperature in the nanorod after being absorbed by the metamaterial, resulting in the change of the dielectric constant of Au, thus changing the transmission of the zero-index cavity [83]. Figure 13(b) shows the simulated transmission spectra of the modulator in ON and OFF states. It can be clearly seen that the transmission of the zero-index cavity can be flexibly tuned by the change of electron temperature $T_e$. When $T_e = 300K$, the modulator is in the OFF state, and the transmission of the zero-index cavity is high, which is marked by the blue line in Fig. 13(b). However, when $T_e = 3000K$ (ON state), the system is deviation from ENZ condition, thus the transmission is low. Importantly, the all-optical modulation overcomes the barrier that the modulation speed of tranditional modulator is slow, thus enables integrated nanoscale switches and modulators in Si waveguides [83]. The optical switching realized by the large nonlinear response of zero-index cavity can be well used in the applications of optical control [200-206].

In addition, zero-index cavity with PEC boundary has been used to realize the geometry-invariant resonant cavities in Fig. 3. Especially, this property can be extended to the open ENZ cavity, in which the switching between radiating and non-radiating modes enables a dynamic control of the emission. The schematic of the ENZ cavity for radiation switching is shown in Fig. 13(c), in which a vacuum spherical bubble, containing a quantum emitter (QE), is attached to a membrane. When the QE is placed in the center $\Delta x = 0$, the cavity mode corresponds to the non radiative mode, that is, the field will be confined in the cavity, as shown in Fig. 13(d). However, when the QE deviates from the center $\Delta x = 3.5\mu m$, the cavity mode corresponds to the radiation mode, and the field radiates strongly to the outside of the cavity, which is shown in Fig. 13(e). Compared with Figs. 13(d) and 13(e), it can be found that the optical switching can be realized when the position of the QE can be changed by the external control such as the vibrational modes excited by an external optical/acoustic wave [83]. Moreover, the interaction between two QEs within an open ENZ cavity is studied in Fig. 13(f). The QE in the left spherical bubble is excited. When the left emitter located in the center of the bubble $\Delta x = 0$, it excites nonradiating mode and two QEs decouple effectively. However, when the position of the QE in the left bubble is separated from such symmetrical positions $\Delta x = 1\mu m$, two QEs will be coupled resonantly. Therefore, this study reveals that switching between radiating mode and nonradiating mode is possible. This provides a new way to control the emission characteristics of QEs, such as enhancing/suppressing the spontaneous emission of QEs [207-209], and dynamically controlling the spontaneous emission of QEs to activate/disable the coupling between them [83].

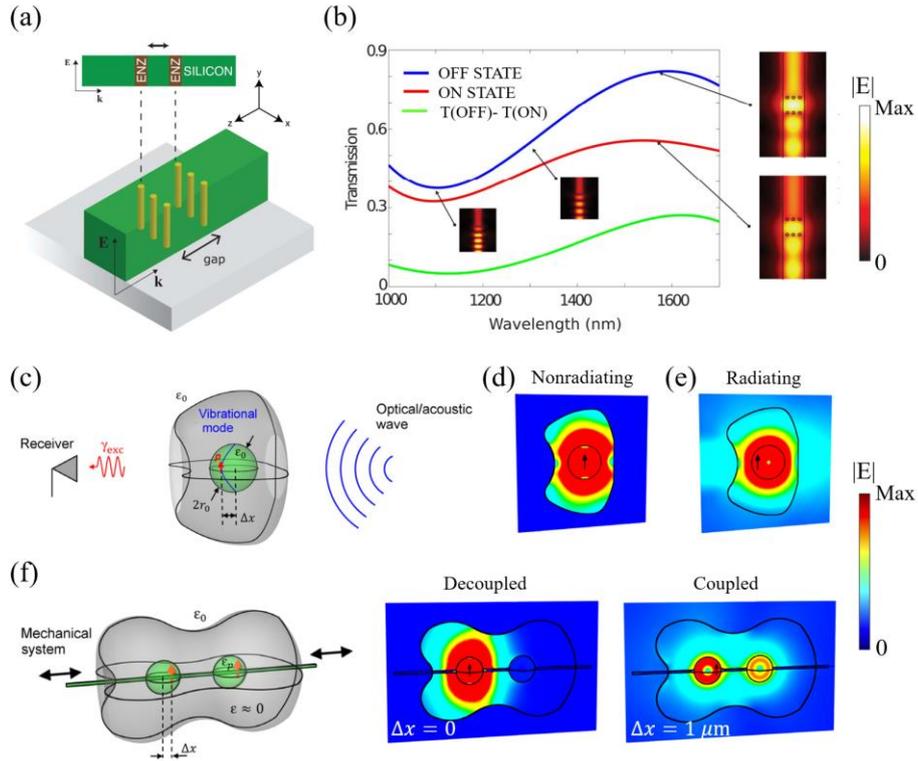

FIG. 13. **Switching realized by the ENZ cavities**. (a) Schematic of the nonlinearity modulator composed of two layers of ENZ media integrated in a conventional Si waveguide. (b) Simulated transmission spectra of the modulator in ON and OFF states. Right insets show the transmission states for the mode propagating from the bottom to the top of the waveguide. Reproduced with permission from Neira et al., Photon. Res. 6, B1 (2018). Copyright 2018, OSA Publishing. (c)-(e) Switching between nonradiating and radiating modes of ENZ cavity. (c) Schematic of the arbitrarily shaped open ENZ cavity, with a vacuum spherical bubble. The position of the inner QE can be displaced by an external stimulus. (d) Simulated electric field distribution for displacement Δx = 0, which corresponds to a nonradiating mode. (e) Similar to (d), but for the displacement Δx = 3.5 μm, which corresponds to a radiating mode. (f) Coupling/decoupling of two QEs in open ENZ cavities. Reproduced with permission from Liberal et al., Sci. Adv. 2, e1600987 (2016). Copyright 2016 Author(s), licensed under a Creative Commons Attribution-NonCommercial-NoDerivatives 4.0 International License.

### 2. Nonreciprocal transmission

In Sec. C.2, the enhanced MO effect have been presented in the 1D heterostructure, which is associated with the tunneling mode of EZI cavity. Recently, the nonreciprocal properties of MO ZIMs have attracted people's great attention [210, 211]. In addition, unpaired Dirac point based on the 2D PCs have been demonstrated can be used to realize the MO ZIM [212]. The nonreciprocal transmission in nonlinear *PT*-symmetric ZIM has been proposed [213]. Here, we will introduce the realization of enhanced nonreciprocal transmission using MO zero-index cavity [82]. Figure 14(a) shows the schematic of a 1-D magnetophotonic crystal with a magnetized ENZ defect. Under an applied magnetic field is in the $y$ direction, the permittivity tensor of ENZ layer is:

$$\tilde{\varepsilon} = \begin{pmatrix} \varepsilon_{xx} & 0 & -i\alpha \\ 0 & \varepsilon_{yy} & 0 \\ i\alpha & 0 & \varepsilon_{zz} \end{pmatrix},$$ (22)

where $\varepsilon_{xx} = \varepsilon_{yy} = \varepsilon_{zz} = \varepsilon_c$ is the diagonal component of permittivity tensor. For such a magnetized medium, TM and TE modes ($H$ or $E$ fields are polarized along the $y$ direction, respectively) are completely decoupled. The field intendity can be significantly enhanced by the ENZ media, thus the MO effect of ENZ cavity mode is stronger than that of the normal cavity mode. The field enhancement as a function of the layer thickness and the incident angle is shown in Fig. 14 (b). When the ENZ layer is embed into a 1-D PC, combining the field enhancement effect of the ENZ layer with the confinement effect of the PC barrier, the MO effect will be further enhanced. Based on the transfer matrix method, the transmission spectrum of the 1D PC with MO zero-index cavity ($\varepsilon_c = 0.1, \alpha = 0.06$) is shown in Fig. 14(c). The wavelengths of defect modes are different for the forward and backward incidences when the light is obliquely incident ($\theta = 30^o$) on the structure. For forward propagation, the transmitted peak appears at 690 nm (red solid line). However, for backward propagation, the transmitted peak appears at 707 nm (blue dotted line). Thus the wavelength of transmitted peak of backward propagation locates at the bandgap of the forward propagation, as shown in Fig. 14(c). This transmission comes from the time-reversal symmetry breaking and a nonreciprocal transmission occurs as the magnetized ENZ defect is introduced into the 1-D PC. As a comparison, figure 14(d) shows the transmission spectrum of using bismuth iron garnet defect ($\varepsilon_c = 6.25, \alpha = 0.06$) with the same MO coefficient instead of the MO ENZ media. It can be clearly seen that the transmitted peaks of the forward (red solid line) and the backward propagations (blue dotted line) nearly overlap at a wavelength of 633.5 nm. The results of MO zero-index cavity in Fig. 14(c) can be validated for enhanced nonreciprocal transmission by simulating the propagation behaviors of electromagnetic waves at wavelength 707 nm. For the forward propagation, we can see from Fig. 14(e) that the input light (marked by cyan arrow) will be totally reflected (marked by the green arrow). However, there is a significant change in the case of backward propagation. The input light will tunnel through the structure (marked by the yellow arrow) with a high transmission, as is illustrated in Fig. 14(f). Therefore, the zero-index cavity can significant enhancement of MO effect and relize the obvious nonreciprocal transmission.

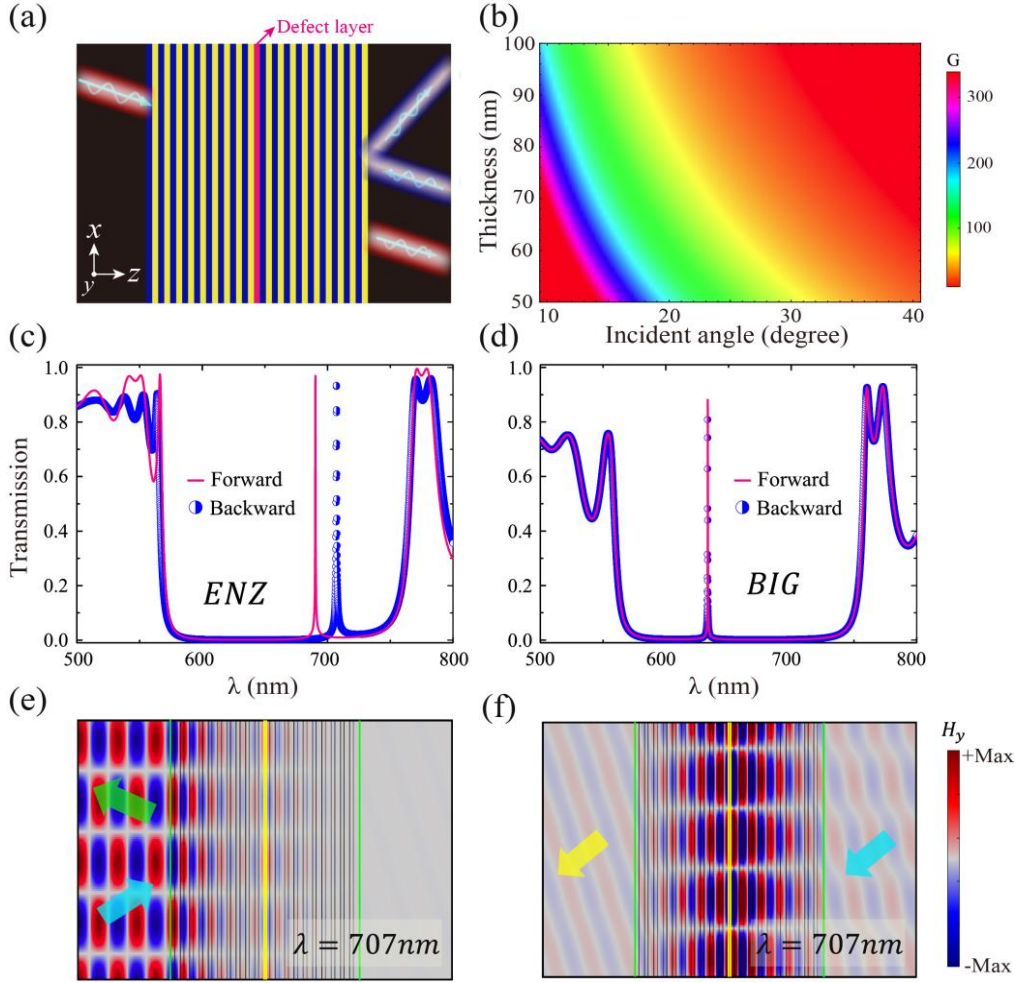

FIG. 14. **Unidirectional transmission realized by the ENZ cavities**. (a) Schematic of a 1-D magnetophotonic crystal with a magnetized defect (marked by the pink layer). (b) The enhancement of field intensity as a function of the incident angle and thickness of layer when a TM-polarized plane wave is incident on an ENZ layer at 400 THz. (c) Calculated transmission of forward (the red trace) and backward propagations (the blue trace) for the 1D PC with a magnetized ENZ defect. (d) Similar to (c), but for a BIC defect. (e) (g) Magnetic-field distributions of the 1-D PC with a magnetized ENZ defect at λ = 707 nm. Reproduced with permission from Guo et al., J. Appl. Phys. 124, 103104 (2018). Copyright 2018, AIP Publishing.

### 3. Collective coupling

The interaction between the cavity field and atoms plays an important role in the CQED. Generally, in order to ensure the strong coupling between the quantum emitter and photons, it is necessary to place the QE at the maximum of the cavity field, which is very challenging in the manufacturing process. Therefore, due to the limited choice of quantum dots and the position uncertainty caused by the inhomogeneity of cavity field, it is difficult to improve the coupling strength between quantum dots and cavity field. In 2012, Jiang *et al.*, proposed theoretically and verified experimentally that the zero-index cavity with uniformed fields can be used to overcome this limitation [76]. The position-independent normal-mode splitting in cavities filled with ZIMs is demonstrated in the circuit-based system, in which the oscillator is constructed by the metallic SRR. Figures 15(a)-15(d) show four samples that an EZI cavity coupled by a SRR at four different places, respectively. The corresponding transmission spectrum of four samples are simulated and measured in Figs. 15(e) and 15(f), respectively. It can be clearly seen that the two split modes are nearly invariant when the SRR is put at four different places of the zero-index cavity. The measured results that in general agree well with the simulations demonstrate a nearly position-independent mode splitting. Therefore, owing to the uniform cavity fields, position-independent normal-mode splitting has been achieved, thus zero-index cavity provide us a new platform to study the interactions of atoms and photons.

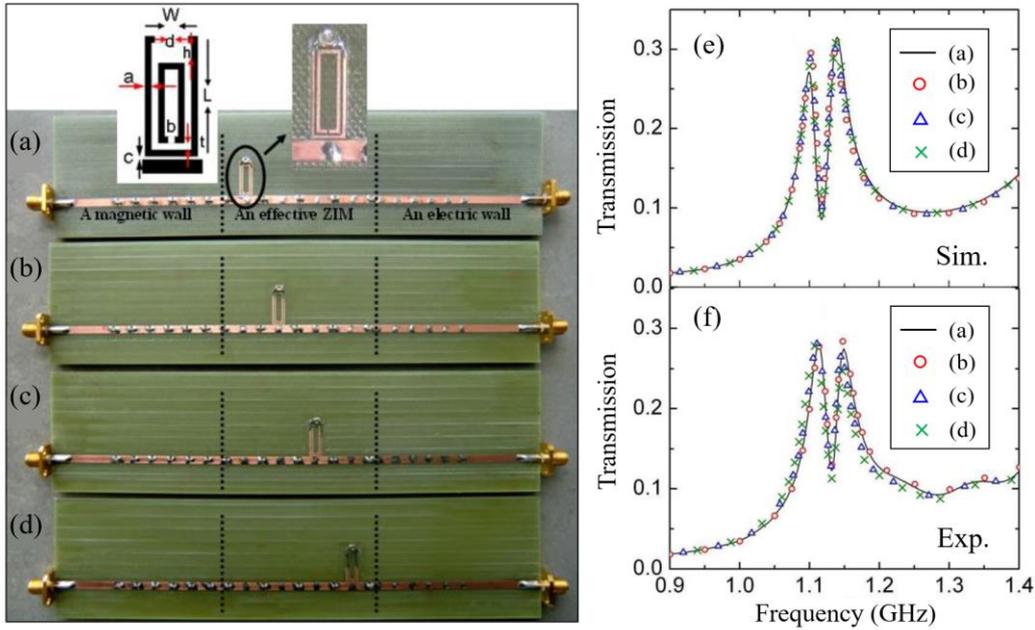

FIG. 15. **Obervation of the position-independent normal-mode splitting in ZIM cavity**. (a)-(d) Photograph of an EZI cavity coupled by a SRR at four different places, respectively. Simulated (e) and measured (f) transmission when a SRR is coupled to an EZI cavity at various locations in (a)-(d). Reproduced with permission from Jiang et al., Opt. Express 20, 6348 (2012). Copyright 2012, OSA Publishing.

The position-independent mode splitting can be extended to study the collective coupling of randomly dispersed oscillators with zero-index cavity [77]. When $N$ oscillators with $G = G_0$ are randomly assigned to a zero-index cavity, due to the enhancement of uniform field, the maximum coupling field of each oscillator is obtained. $N$ oscillators are equivalent to an effective $NG_0$ oscillator and the collective coupling is realized. Without the loss of generality, figure. 16 shows the collective coupling in 1D PC with zero-index cavity. The schematic of a 1D PC with normal cavity [(AB)₅A(AB)₅] in $x$-$z$ plane is presented, in which A and B denote SiO₂ and Ta₂O₅ with $\varepsilon_A = 2.13$, $\varepsilon_B = 4.53$, respectively. The thickness of layers A and B is same $d_A = d_B = 100$ nm. The corresponding transmission spectrum is shown in Fig. 16(b). The cavity mode ($f = 433$ THz) with exist in the band gap is marked by the red dashed line. In Fig. 16(c), the field of the cavity mode is mainly localized at the cavity region and it decays exponentially with the position away from the cavity. The schematic of a zero-index cavity embedded in 1D PC is shown in Fig. 16(d). It is found that the frequency of cavity mode ($f = 433$ THz) does not change after inserting the ZIM layer into the structure in Fig. 16(a). The corresponding field distribution of the cavity mode in the zero-index cavity is shown in Fig. 16(e). It is seen that the enhanced electric fields are uniform in the zero-index cavity and conforms to Maxwell's equation and boundary conditions. When considering placing multiple oscillators [marked by the white crosses in Fig. 16(e)] into a zero-index cavity, the collective coupling will be verified, as shown in Fig. 16(f). The resonant susceptibility of the oscillator is described by [77, 214]:

$$\chi_r = -G_0 / [(\omega - \omega_0) + i\gamma], \tag{23}$$

where $\omega_0 = 2.72$ PHz. The oscillating strength $G_0$ and damping factor $\gamma$ are assumed as 0.14 PHz and 2.51 THz, respectively. From the calculated spectra of mode splitting for the zero-index cavity with one ($N = 1$), two ($N = 2$), three ($N = 3$) and four ($N = 4$) oscillators, it is seen that the splitting interval of frequency $\Delta\Omega$ enlarges when the number of oscillators increases, which is shown in Fig. 16(f). Especially, the mode splitting is proportional to $\sqrt{NG}$ and the splitting interval of frequency $\Delta\Omega$ for $N$ oscillators with $G = G_0$ is equal to that for one oscillator with $G = NG_0$. Therefore, one can tune the collective coupling between the oscillator can cavity field by varying the number of oscillators in the zero-index cavity. Moreover, the collective coupling has also been studied in a practical system that the ZIM cavity is realized by a 2D dielectric PC at Dirac-like point [147]. The schematic of a 1D PC cavity embedded with 2D PC-based ZIM is shown in Fig. 16(g). Similar to Figs. 16(e) and 16(f), the field distribution and splitting interval spectrum of the practical zero-index cavity for collective coupling are shown in Figs. 16(h) and 16(i), respectively. The four points in Fig. 16(f) also have linear dependency, which reveals that $\Delta\Omega \propto \sqrt{N}$.

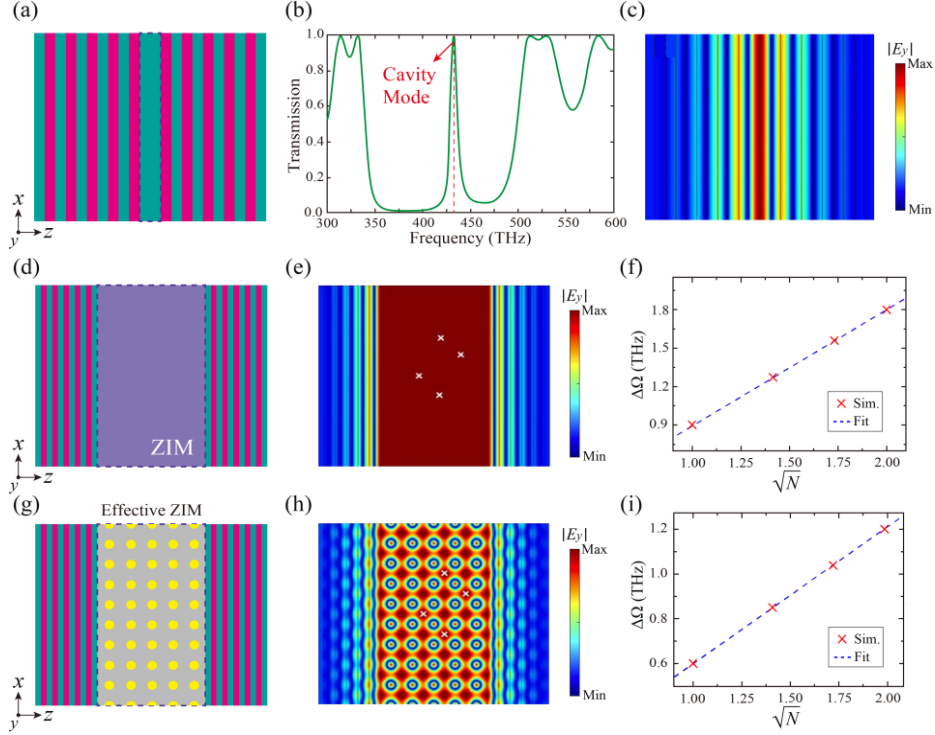

FIG. 16. **Collective coupling of randomly dispersed resonators in the ZIM cavity based on 1D PCs**. (a)-(c) Normal cavity mode in 1D PC. (a) Schematic of a 1D all-dielectric PC: (AB)₅A(AB)₅, in which a defect layer A is placed at the center of the structure. (b) Transmission of the structure (AB)₅A(AB)₅. The cavity mode is marked by the dashed red line. (c) The corresponding electric field distribution of the normal cavity mode. (d)-(i) ZIM cavity mode in 1D PC. (d) Schematic of a 1D PC cavity embedded with ZIM. (e) Electric field distribution of the ZIM cavity mode. In the ZIM cavity, four white crosses indicate randomly dispersed resonators. (f) The mode splitting interval versus the square root of $N$ (the number of resonators). (g)-(i) Similar to (d)-(f), but for a 1D PC cavity embedded with 2D PC-based ZIM. Reproduced with permission from Xu et al., Eur. Phys. J. B 87, 44 (2014). Copyright 2014, EDP Sciences, SIF, Springer-Verlag Berlin Heidelberg.

Moreover, the collective coupling realized by the zero-index cavity has been demonstrated in the microwave experiments based on the circuit-based ZIM and SRRs [77]. Figures 17(a)-17(d) show four samples that an EZI cavity coupled by one, two, three and four SRRs, respectively. The corresponding transmission spectrum of four samples are simulated and measured in Figs. 17(e) and 17(f), respectively. It can be clearly seen that the splitting intervals enlarge as the number of SRRs increases. The simulation results are in good agreement with the experimental results. Especially, the splitting intervals as a function of the square root of $N$ is shown in Fig. 17(g). There is a linear correlation between the simulated points and the measured points near the blue line. The small deviation is due to the influence of the losses in the samples. The collective coupling realized by the zero-index cavity may be utilized to alleviate the limited choice of quantum dots and the positional uncertainty brought by the inhomogeneous cavity fields [77].

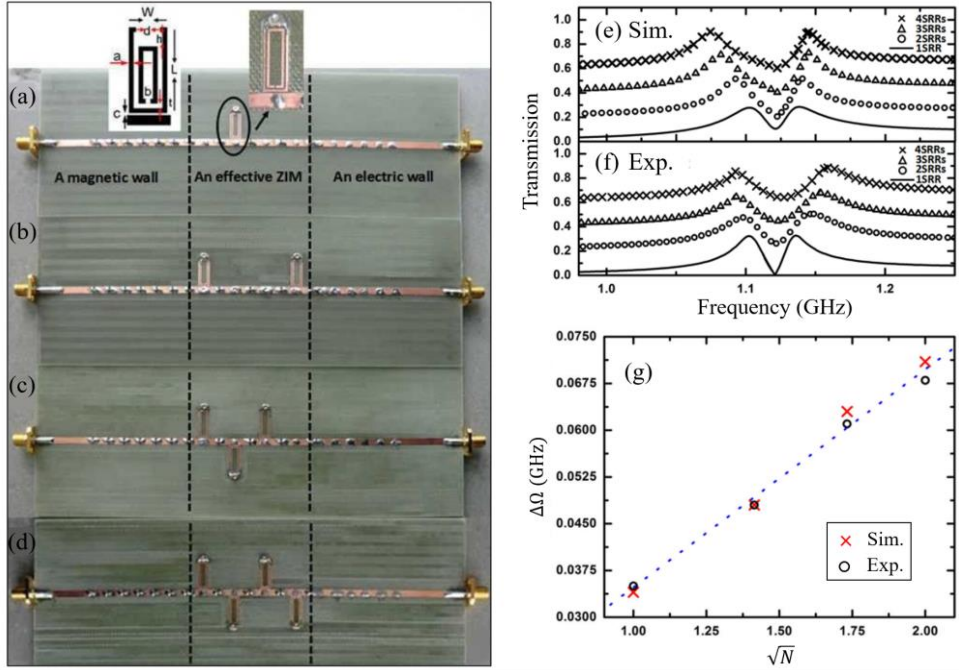

FIG. 17. **Obervation of the collective coupling in the ZIM cavity**. Photograph of an EZI cavity coupled by: (a) one, (b) two, (c) three, and (d) four SRRs. Simulated (e) and measured (f) transmission when different numver of SRRs are coupled to an EZI cavity in (a)-(d). (g) The simulated and the measured splitting intervals derived from (e) and (f) versus the square root of $N$ (the number of SRRs), as indicated by the red crosses and the open circles, respectively. Reproduced with permission from Xu et al., Eur. Phys. J. B 87, 44 (2014). Copyright 2014, EDP Sciences, SIF, Springer-Verlag Berlin Heidelberg.

In Sec. II, we systematically introduce the realization and physical properties of the zero-index cavity. Moreover, the EZI cavity with zero thickness is presented based on the tunneling mode in the heterostructure. The zero-index cavity not only enhance the interaction between light and matter to increase the nonlinear and MO response, but also realize the position independent coupling and collective coupling effect. Especially, the actively controlled ZIMs, such as electronic control [215], magnetic control [216] and optical pumping [217], greatly enrich the design flexibility of zero-index cavities. In addition, the low-loss ZIMs based on bound states in the continuum (BIC) have been proposed [218-223], whose physical mechanism is mainly to suppress the out of plane radiation at the Dirac-like point [221-223]. Therefore, the ZIMs and their designed zero-index cavities are gradually attracting people's attention. In addition to the important applications described in this section, they may also be used in many aspects, including absorber [224-226], shielding [157, 227], splitter [228], sensor [229, 230], isolator [231], antenna [232, 233], and so on.

# III. HYPERBOLIC METACAVITES

The interaction between light and matter depends on media dispersion in momentum space, which can be characterized using IFCs. The topological transition of dispersion from a closed sphere in isotropic media to an open hyperboloid in anisotropic media has been achieved by various means [234-238]. Especially, HMMs with hyperbolic dispersion exhibit many intriguing features and attracted people's great attention. One of the most important characteristics of HMM is that it can control the near field effectively. Because of the open IFCs, HMMs support propagating EM waves with large wave vectors [239-241]. The capabilities of HMMs have been demonstrated in numerous applications that utilize their exotic high-$k$ modes, such as enhanced spontaneous emission [242-248], long-range interactions [249-253], superresolution imaging [254-257], optical pulling forces [258, 259], and high-sensitivity sensors [260-268]. For electric/magnetic HMM, the principal components of its electric/magnetic tensor have opposite signs. Based on the Maxwell's equations, the dispersion relation of an electric anisotropic material ($\mu_x = \mu_y = \mu_z = 1$) can be written as [35-40]:

$$(k_x^2 + k_y^2 + k_z^2 - \varepsilon_{//} k_0^2)[(k_x^2 + k_y^2)/\varepsilon_{\perp} + k_z^2/\varepsilon_{//} - k_0^2] = 0 , \qquad (24)$$

where $k_x$, $k_y$, and $k_z$ denote the $x$, $y$, and $z$ components of the wavevector, $k_0$ is the wave-vector in free space. $\varepsilon_{//}$ and $\varepsilon_{\perp}$ denote the permittivity components perpendicular and parallel to the $xy$ palne. The first and second terms describe the EM response of the TE and TM polarized waves. Especially, in Figs. 18(a) and 18(b), two types of HMMs with onefold and twofold surfaces for the TM polarized wave correspond to the cases: $\varepsilon_{\perp} > 0$, $\varepsilon_{//} < 0$ (type I HMM) and $\varepsilon_{\perp} < 0$, $\varepsilon_{//} > 0$ (type II HMM), respectively. For comparison, the IFC of the air is added as the green sphere. The EM parameter space of anisotropic media is shown in Figs. 18(c) and 18(d). Take $\mu > 0$ in Fig. 18(c) for example, both $\varepsilon_{//}$ and $\varepsilon_{\perp}$ are positive in the first quadrant, and the IFC of the media is a closed ellipse or circle. Type II HMM and type I HMM media correspond to the second and fourth quafrants, respectively. In addition, the third quadrant ($\varepsilon_{\perp} < 0$, $\varepsilon_{//} < 0$) represents the ENG media, in which the EM wave will exist in the form of evanescent wave in the media [269, 270]. The similar EM parameter distribution of anisotropic media for $\mu < 0$ is shown in Fig. 18(d).

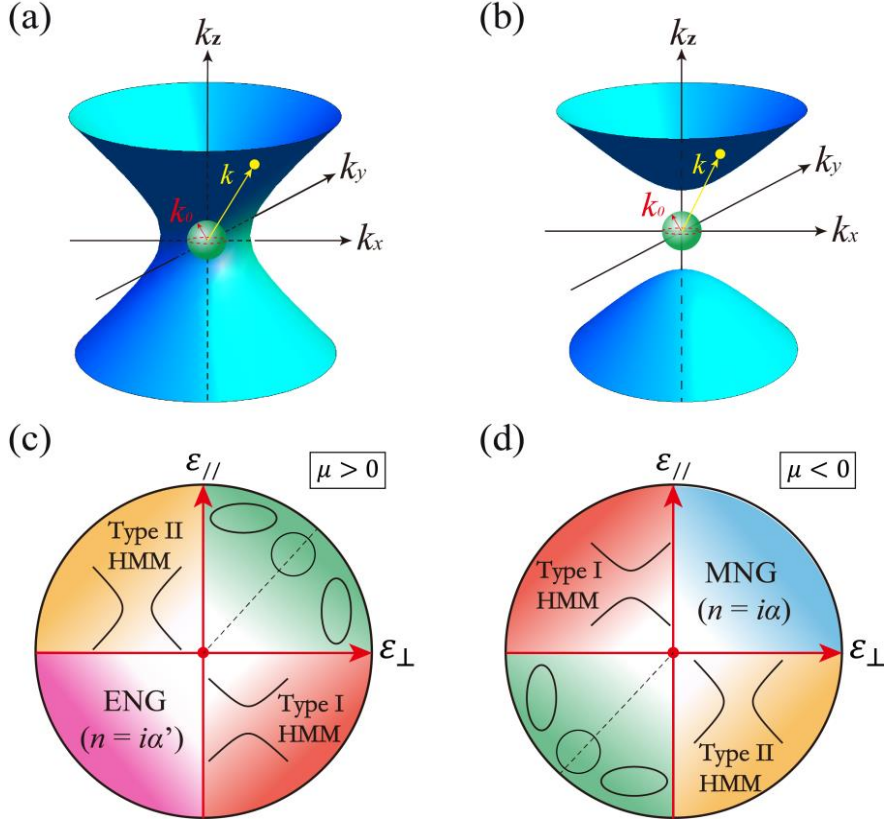

FIG. 18. **Two types of HMMs**. (a) Hyperboloid IFC of metal-type or type II HMM (onefold blue surface) and the spherical IFC of air (green surface). (b) Similar to (a), but for the IFC of dielectric-type or type I HMM (twofold blue surface). (c) Electromagnetic parameter space of the anisotropic media. There are two types electric HMM: type II HMM ($\varepsilon_\perp < 0$, $\varepsilon_{//} > 0$) and type I HMM ($\varepsilon_\perp > 0$, $\varepsilon_{//} < 0$) media for $\mu > 0$. (d) Similar to (c) but for the HMMs with $\mu < 0$.

## A. Realization ways

### 1. Metal/dielectric multilayer hyperbolic media

According to Eq. (1), HMMs with $\varepsilon_\perp \varepsilon_{//} < 0$ can be constructed based on the metal/dielectric multilayers. Recently, the rolled-up HMM also have been proposed base on the metal/dielectric multilayers [271]. The corresponding hyperbolic cavity have been widely studied [93, 94, 102]. Figure 19(a) shows an optical hyperbolic cavity array which is composed of silver/germanium multilayers [93]. Especially, in addition to electric HMM, the magnetic HMM can the associated topological transition can be realized considering the non-local effect of the metal/dielectric multilayer fishnet structure [272, 273]. Figure 19(b) shows the schematic and the sample of the multilayer fishnet structure, which can can be used to relize the negative-index media [274] and HMMs [273]. The 3D optical metamaterials offer the effective avenue to explore a large variety of optical phenomena associated with ZIM, negative-index media, and HMMs.

### 2. Metal nanowire hyperbolic media

Wire metamaterials represent a large class of artificial EM structure, which corresponds the lattices of aligned metal rods embedded in the dielectric host [275]. An important group of subwavelength wire metamaterials possess extreme optical anisotropy based on the EMT:

$$\varepsilon_{//} = \frac{[(1+p)\varepsilon_m + (1-p)\varepsilon_d]\varepsilon_d}{(1-p)\varepsilon_m + (1+p)\varepsilon_d}, \varepsilon_\perp = p\varepsilon_m + (1-p)\varepsilon_d, \tag{25}$$

where the filling ratio of the metal wires is equal to the radio of cross-sectional areas of the metal wires and the host dielectric $p = A/A_0$. $\varepsilon_d$ and $\varepsilon_m$ denote the permittivity of the metal and dielectric, respectively. Especially, the electric HMMs can be realized for $\varepsilon_\perp \varepsilon_{//} < 0$. Figure 9(c) shows one sample of the electric HMM in the visible range, which is constructed by an assembly of Au nanorods electrochemically grown into a substrate-supported, thin-film porous aluminium oxide template [260]. The guided mode in HMM with strong field localization is quite similar to the surface plasmon mode of a solid metal film, thus the nanowire hyperbolic media will provide very high sensitivity to refractive index changes [260]. In addition, similar to the metal/dielectric multilayers, metal nanowire structure not only can be used to design electric HMM, but also can realize the magnetic HMM associated with magnetic topological transition of IFC [276].

### 3. Circuit-based hyperbolic media

In the 2D TL system, circuit-based metamaterials can realize various electromagnetic parameters under EMT [163, 164]. Recently, HMMs with flexbile EM parameters can be well constructed using circuit-based metamaterials in the microwave regime [277-279]. For example, the interesting epsilon-near-pole HMM can be easily constructed base on the circuit-based metmaterials [253]. Figure 19(d) shows a circuit-based HMM sample, in which the real part of anisotropic permeability of the system can be tuned by changing the lumped elements [277]. In addition to the real part, the imaginary part of the EM parameters can be directly controlled by adding the lumped resistors [166, 280]. The actively controlled HMM has been proposed by using variable capacitance diodes under appling external voltage [137]. Importantly, the hyperbolic topological transition and the novel linear-crossing metamaterials (LCMMs) have been proposed and demonstrated in the circuit-based system [281]. So far, circuit-based HMMs have attracted much attention in terms of various applications: emission pattern control [137], long-range atom-atom interaction [253], collimation [280], super-resolution imaging [281], spin-Hall effect [282], etc.

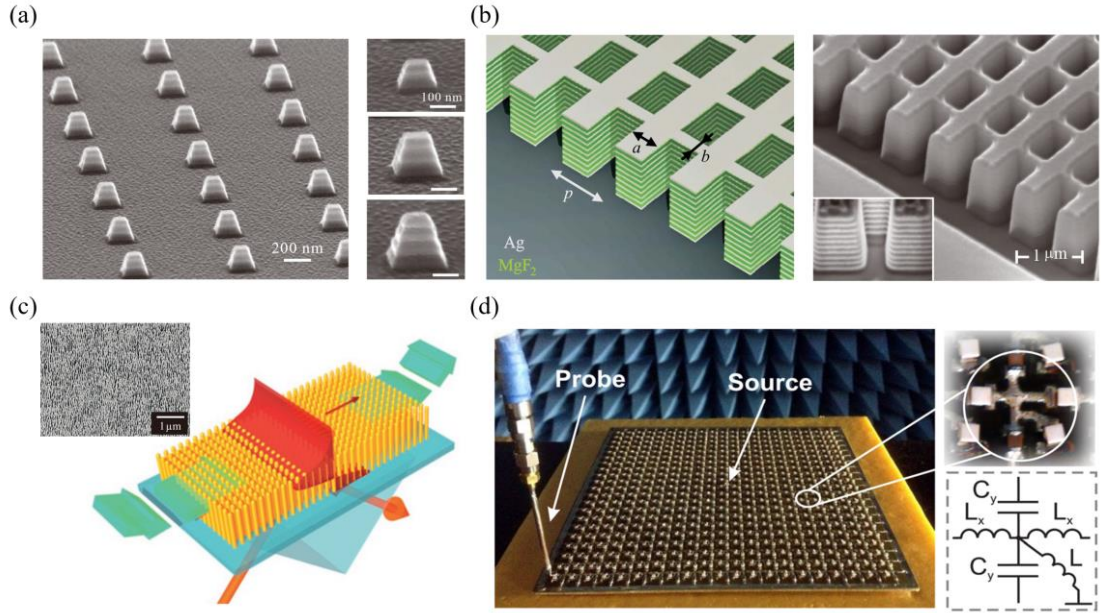

FIG. 19. **Realization of HMMs**. (a) Hyperbolic cavity array, where the HMM is realized by the metal/dielectric multilayer. Insets show SEM images of three cavities with different dimensions. Reproduced with permission from Yang et al., Nat. Photon. 6, 450 (2012). Copyright 2012 Springer Nature. (b) Multilayer fishnet metamaterials. This structure can be used to relize the negative-index metamaterial and HMMs. Inset shows SEM image of the fishnet structure with the side etched, showing the cross-section. Reproduced with permission from Valentine et al., Nature 455, 376 (2008). Copyright 2008 Springer Nature. (c) Plasmonic nanorod metamaterial. This structure can be used to relize the HMMs. Inset shows SEM image of nanorod arrays. Reproduced with permission from Kabashin et al., Nat. Mater. 8, 867 (2009). Copyright 2009 Springer Nature. (d) Circuit-based HMM in the microwave and RF regime. Inset shows enlarged circuit elements and the corresponding effective circuit model. Reproduced with permission from Chshelokova et al., J. Appl. Phys. 112, 073116 (2012). Copyright 2012, AIP Publishing.

## B. Physical properties

### 1. Anomalous scaling law

Because of the special HMM dispersion, the propagation direction of a wave in a HMM is different from that in a normal anisotropic material with a closed IFC. For the traditional optical cavity made of dielectric, whose IFC is a closed circle and the supported wavevector is limited. In addition, the isotropic property of dielectric making the tranditional optical cavity is directional independent, as shown in Figs. 20(a) and 20(b). The dashed lines denote the IFC at a frequency slightly above the frequency of the solid lines, which indicating the gradient direction of the IFCs of the dielectric [40]. As we all know that the wavevector increase with the frequency increase because of their linear relationship $k = n\omega/c$. Specially, when the wavevector in the $x$ direction is fixed (marked by a blue dotted line along the $z$ direction), the wavevector at a higher frequency is larger than one at a lower frequency $k_{z2} > k_{z1}$, as shown in Fig. 20(a). This positive correlation between wavevector and frequency is marked by the sign '+' for see. Similarly, when the wavevector in the $z$ direction is fixed (which is marked by a blue dotted line along the x direction), the wavevector at a higher frequency is also larger than one at a lower frequency in Fig. 20(b). However, the wavevector property of the HMM is significant different along $x$ and $z$ directions, which is presented in Figs. 20(c) and 20(d). When the wavevector in the $x$ ($z$) direction is fixed, the wavevector at a higher frequency is larger (smaller) than one at a lower frequency $k_{z2} > k_{z1}$ ($k_{x2} > k_{x1}$). The abnormal wavevector properties of the HMM is marked '-' in Fig. 20(d). Therefore, it can be expected that the abnormal cavity property appears for the HMM in $x$ direction.

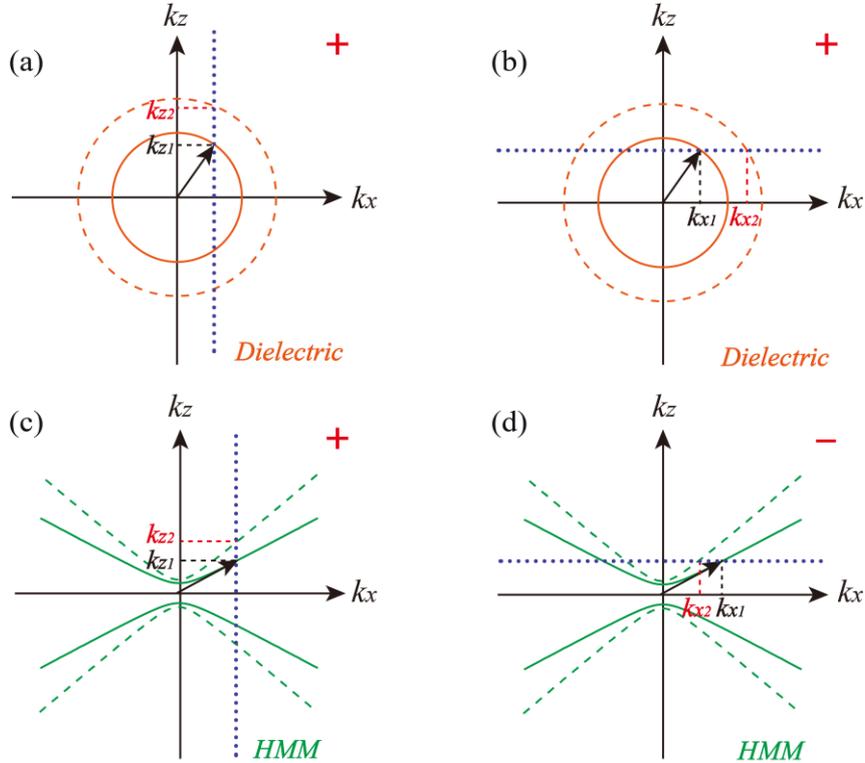

FIG. 20. **Analysis of the anomalous scaling law of hyperbolic cavity with IFCs**. IFCs for the dielectric (a), (b) and HMM (c), (d), where the solid and dashed lines correspond to the values at frequencies ω and ω + δω, respectively. Dotted line in each graph is used to represent the determined $k_x$ or $k_z$. Reproduced with permission from Wang et al., Phys. Rev. Appl. 13, 044024 (2020). Copyright 2020, American Physical Society.

For tranditional optical cavities composed of dielectric, the frequency increases with the increase of mode order due to the standing wave condition. However, this scaling law of the cavity mode will be modified in the HMMs. Especially, the anisotropic scaling law have been demonstrated in the hyperbolic cavity based on metal/dielectric multilayers [93]. The schematic of the hyperbolic cavity consists of alternating thin layers of silver and germanium is shown in Fig. 21(a). The hyperbolic cavity modes were excited with a TM polarized plane wave propagating along the *z*-direction. Especially, the first five lowest order modes along *z* direction are studied, which are labeled (1, 1, 1), (1, 1, 2), (1, 1, 3), (1, 1, 4), and (1, 1, 5), respectively. The corresponding simulated electric field $E_z$ distributions of these modes are shown in Fig. 21; here, it can clearly be seen that the higher-order mode is found at a lower resonant frequency in the hyperbolic cavity. For example, the frequency of the high-order mode (1, 1, 5) is 49.6 THz, while the frequency of the low-order mode (1, 1, 1) is 145.2 THz. By comparing five cavity modes with different order, the anomalous property of the HMM mode order in the *z* direction is demonstrated in Figs. 21(b)-21(f).

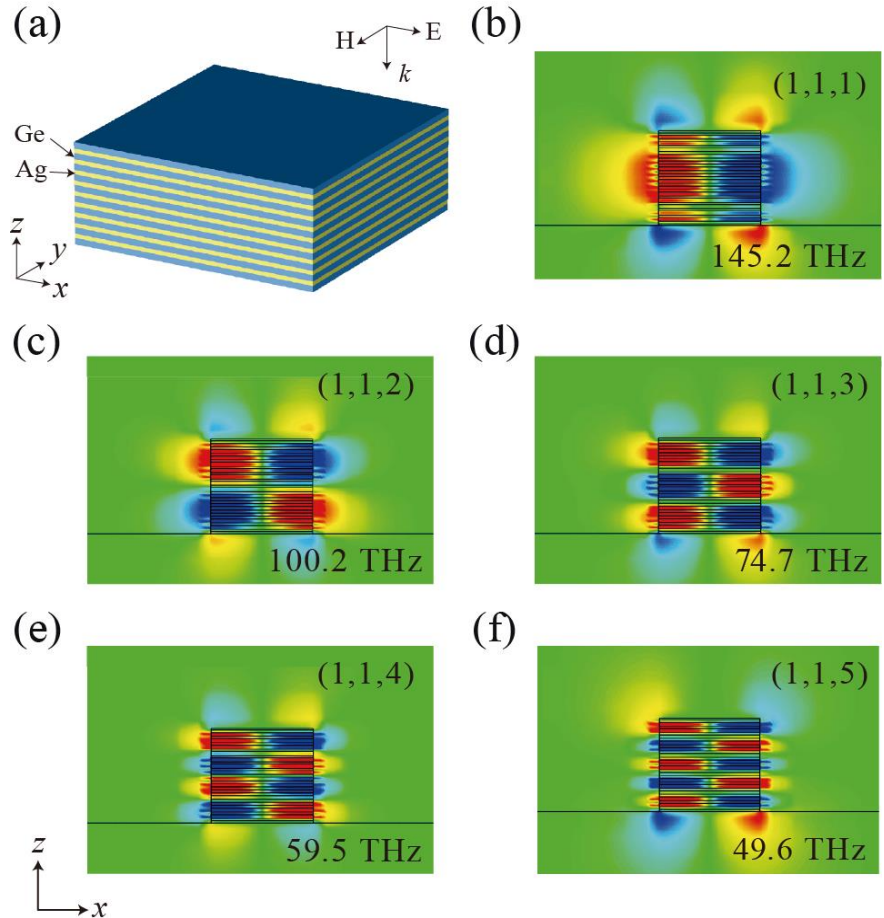

FIG. 21. **Anomalous scaling law in the metal/dielectric multilayer hyperbolic cavity.** (a) Schematic of Ag/Ge multilayer hyperbolic cavity. (b)-(f) Simulated $E_z$ distribution of the first five hyperbolic cavity modes along the *z*-direction. Reproduced with permission from Yang et al., Nat. Photon. 6, 450 (2012). Copyright 2012 Springer Nature.

The anomalous scaling law of hyperbolic cavity also have been clearly demonstrated based on the circuit-based hyperbolic cavity in the microwave regime [109]. The schematic of the near-field detection system is shown in Fig. 22(a). A subminiature version A (SMA) connector that functions as the source for the system is placed at the center of the sample as a vertical monopole to excite the circuit-based prototype. A small homemade rod antenna is employed to measure the out-of-plane electric field at a fixed height of 1 mm from the planar microstrip. Figure 22(b) shows the measured normalized $E_y$ spectrum for the circuit-based hyperbolic cavity. Four cavity modes with different order: $C_{71}$, $C_{51}$, $C_{31}$, and $C_{11}$ are marked by the red arrows. The corresponding measured $E_y$ distributions of the cavity modes are shown in Figs. 22(c)–22(f), respectively. Similar to the hyperbolic cavity realized by the metal/dielectric multilayers in Fig. 21, the experimental results of the circuit-based hyperbolic cavity show that the higher order mode is found at a lower resonant frequency, thus the anomalous mode-order property of the hyperbolic cavity in the $x$ direction is experimentally demonstrated in the microwave regime [109].

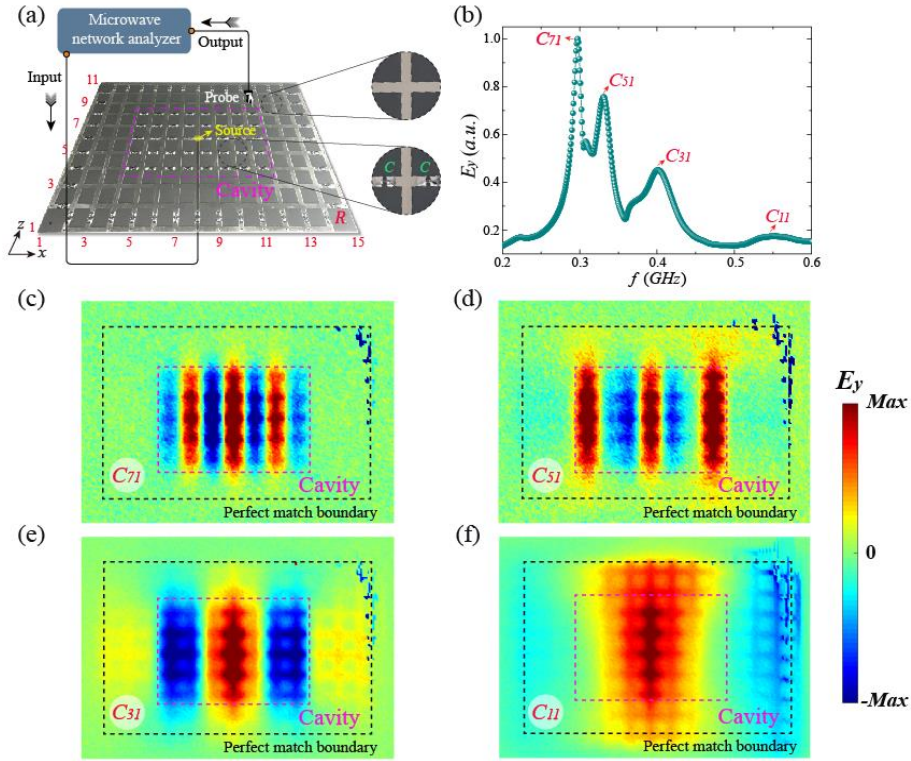

FIG. 22. **Observation of anomalous scaling law in the circuit-based hyperbolic cavity.** (a) Experimental schematic of the TL-based hyperbolic cavity. Insets show the amplified real structure photos of HMM and background DPS medium, respectively. (b) Measured normalized $E_y$ spectra for the hyperbolic cavity modes. First four hyperbolic cavity modes are marked by different lables. (c)–(f) Measured near-field distributions of $E_y$ at four different frequencies: (c) 0.295 GHz (order $C_{71}$), (d) 0.331 GHz (order $C_{51}$), (e) 0.401 GHz (order $C_{31}$), and (f) 0.550 GHz (order $C_{11}$). Reproduced with permission from Wang et al., Phys. Rev. Appl. 13, 044024 (2020). Copyright 2020, American Physical Society.

### *2. Size independent cavity mode*

Hyperbolic cavity exhibits the abnormal scale property along the direction that perpendicular to the optial axis of the HMM. As introduced in Fig. 20 that the abnormal (normal) cavity property appears for the HMM in $x(z)$ direction. Therefore, a question naturally arises: can the size independent cavity mode be realized in hyperbolic cavity by changing the different length in different directions? The dependence of the hyperbolic cavity mode on the structure size in the metal/dielectric multilayer structure (same as the structure in Fig. 21) is studied in Fig. 23. The simulated IFC of the effective HMM for TM polarized wave is obtained from the Fourier spectrum in Fig. 23(a). Five cavities with different size combinations (width, height) at 150 THz are taken as examples. The corresponding simulated $E_y$ distributions of hyperbolic cavities with different size are shown Fig. 23(b). It is seen that five cavities with different size support identical optical modes with the same resonant frequency (150 THz) and the same mode order (1, 1, 1) [93]. This property has also been experimentally demonstrated from the transmission spectra in Fig. 23(c). It can be clearly seen that the hyperolic cavity with different size resonate at the same resonant frequency 205.5 THz (147 THz) for the left (right) panel. The simulated total radiation quality factor $Q_{rad}$ of the cavity modes follows the fourth power law between $Q_{rad}$ and $k$ as $Q_{rad} \sim (k/k_0)^4$, which is shown in Fig. 23(d). Moreover, the vertical coupling radiation quality factors between the plane wave and the cavity mode also follows the fourth power law $Q_{rad,v} \sim (k/k_0)^4$ for different height, width, mode order and resonant frequency, as shown in Fig. 23(e). Especially, cavity area filling ratios of 5% and 10% are considered. The measured results are meet well with the theoretical predicted results. Therefore, hyperbolic cavity provides an effective way to increase the light-matter interaction, thus it may open up new possibilities for various applications such as CQEQ, optical nonlinearities, and biosensing [93].

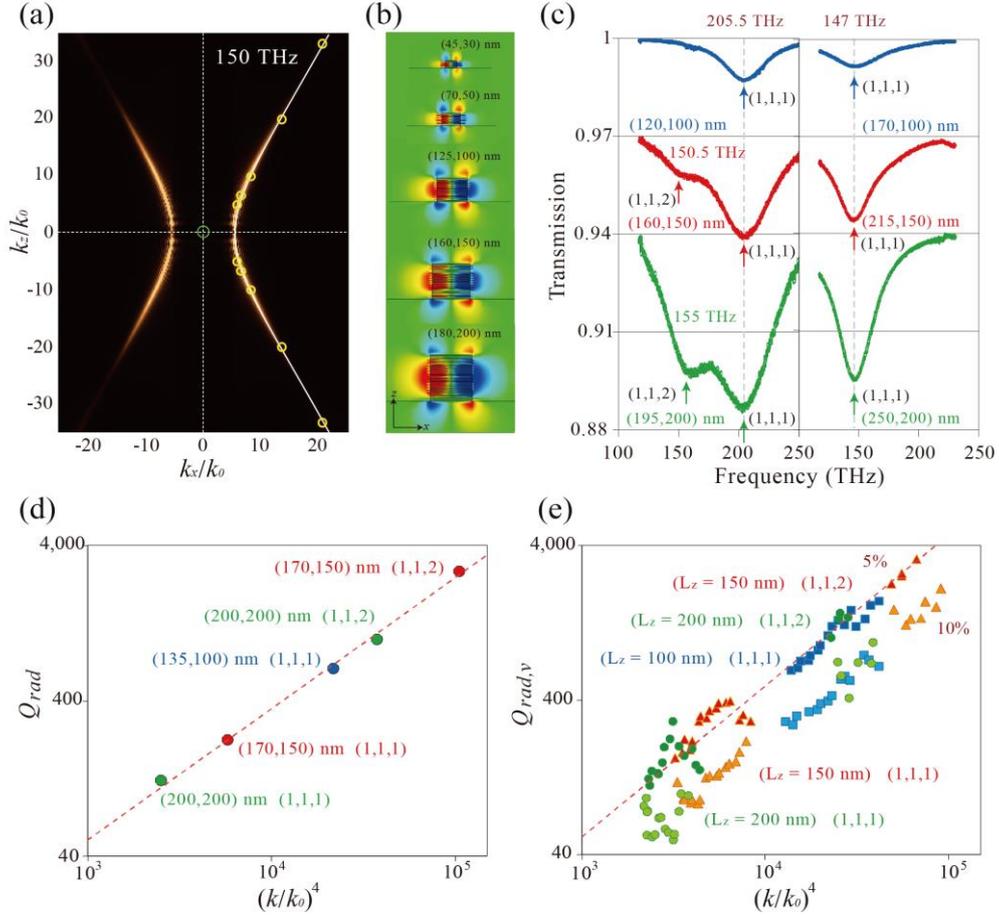

FIG. 23. **Observation of size independent cavity mode in the metal/dielectric multilayer hyperbolic cavity.** (a) FDTD-simulated IFC of the multilayer HMM at 150 THz. (b) Simulated $E_z$ distributions of the (1,1,1) mode for hyperbolic cavities with different size (width, height) combinations but at the same resonant frequency of 150 THz. (c) Measured transmission spectra of hyperbolic cavities with different sizes and different frequencies. (d) Simulated total radiation quality factor $Q_{rad}$ of a single hyperbolic cavity scales as $(k/k_0)^4$ or $n_{eff}^4$, a universal fourth power law, for cavities with different dimensions, resonance frequencies and mode orders. (e) The retrieved vertical radiation quality factor $Q_{rad,v}$ shows the same power law. Reproduced with permission from Yang et al., Nat. Photon. 6, 450 (2012). Copyright 2012 Springer Nature.

The dependence of the hyperbolic cavity mode on the structure size is also demonsrated based on the circuit-based hyperbolic cavity for TE polarized wave [109]. The size of the hyperbolic cavity is identified by the number of unit cells in the $x$ and $z$ directions. The schematic of the circuit-based hyperbolic cavity with $N_x = 7$ and $N_y = 5$ is shown in Fig. 24(a). In fact, the realization of size independent cavity mode in hyperbolic cavity is come from the anisotropic scaling law. Especially, the effect of changing the length of hyperbolic cavity in $x$ and $z$ directions on the cavity mode is presented in Figs. 24(b) and 24(c), respectively. The cavity mode $C_{11}$ in different cavities are marked by red circles for visualization. It can be clearly seen that the frequency of mode $C_{11}$ blueshifts (redshifts) with decreasing length in the $x(z)$ direction. Because the dependence of the frequency of cavity mode on the structure's size is opposite to that in the $x$ and $z$ directions, the hyperbolic cavities with different size resonant at same frequency can be realized. Figure 24(d) shows the measured $E_y$ spectrum for two circuit-based hyperbolic cavities: $N_x = 7$, $N_y = 5$ and $N_x = 4$, $N_y = 4$. While the size of the cavity with $N_x = 7$, $N_y = 5$ is significantly larger than that of the cavity with $N_x = 4$, $N_y = 4$, the frequency of cavity mode $C_{11}$ is the same for both cavities. Therefore, the size independent cavity mode is experimentally demonstrated in the circuit-based hyperbolic cavity. Especially, the circuit-based hyperbolic cavities not only extend previous research work on hyperbolic cavities to magnetic HMMs, but they also have a planar structure that is easier to integrate and has a smaller loss. The hyperbolic cavities may enable their use in some microwave-related applications, such as in high-sensitivity sensors and miniaturized narrowband filters [109].

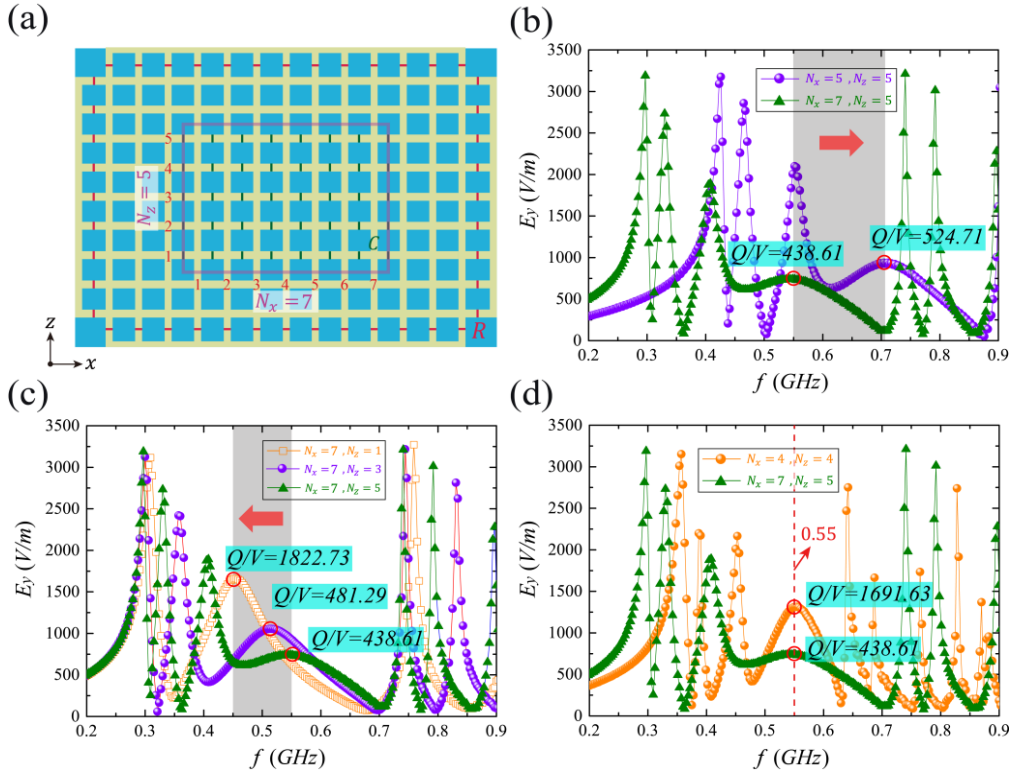

FIG. 24. **Observation of size independent cavity mode in the circuit-based hyperbolic cavity.** (a) Schematic of the circuit-based hyperbolic cavity with $N_x = 7$ and $N_y = 5$. (b) Simulated $E_y$ spectra for the hyperbolic cavities with different $N_x$, and the cavity modes blueshift with decreasing $N_x$. (c) Similar to (b), but for different $N_z$, and the cavity modes red-shift with decreasing $N_z$. (d) Simulated $E_y$ spectra for $N_x = 4$, $N_z = 4$ and the frequency of the $C_{11}$ mode is the same as that of the $C_{11}$ mode in the hyperbolic cavity with $N_x = 7$, $N_z = 5$. 

### 3. Continuum of high-order modes

For conventional cavity mode, any mode order appears at one single frequency. It is because that when the incident angle of the light is fixed, the wavevector in the propagating direction is given. However, once the special wavevector in the propagating direction corresponds to more than one frequency, the mode continuum allowing multiple manifestations of the same order can be realized [103]. The specialy dispersion of HMM not only enables the novel resonant modes with anomalous scaling law [92, 93, 109] but also may be used to form a mode continuum [103]. The schematic of special hyperbolic cavity for mode continuum is composed of a core layer of HMM, consisting of metal/dielectric multilayers, and two cladding mirrors, as shown in Fig. 25(a). In this case, the the wave-vector component $k_z$ solely defines the resonance condition, and thus determines the mode spectrum. The $k_z$ of light with incident angle $\theta$ in the HMM core can be written as [103]:

$$k_z(\omega, \theta) = \frac{\omega}{c} \sqrt{\varepsilon_\perp(\omega) \left(1 - \frac{\sin^2 \theta}{\varepsilon_{//}(\omega)}\right)}. \tag{26}$$

Interestingly, $k_z$ is nonmonotonic with angular frequency $\omega$ at a fixed $\theta$ in the HMM. As a result, the mode continuum can be realized in the hyperbolic cavity. Figure 25(b) shows the dispersion relation of the microcavity for TM polarization. There are some interesting properties can be found. (1) There are two curves correspond to the cavity mode of the second order. Especially, the slope of the lower frequency second-order mode is negative, which corresponds to a anomalous cavity mode with negative group velocity. Therefore, this configuration gives rise to modes of identical orders appearing at different frequencies. (2) A novel unconventional zeroth-order mode based on the phase compensation appearing at $k_x = 1.08 \times 10^6 m^{-1}$. (3) Between the anomalous second- and zeroth-order modes, a cluster of high-order modes exist, forming a mode continuum [103]. The continuum of high-order modes is enlarged at the inset of Fig. 25(b). It is seen that the mode order increase with the increase of $k_x$. The anomalous resonant modes of Hyperbolic cavity are demonstrated from the reflection spectrum in Figs. 25(c) and 25(d). For TE polarized wave, the structure of metal/dielecric multilayers corresponds a dielectric instead of the HMM. As a result, reflection spectra correspond to those of a conventional microcavity with positive dispersion, as shown in Fig. 25(c). The cavity modes are indicated by the reflection minimas in the refelction spectrum, which are marked by the dashed lines. Howeve, when the incident agnle larger than $\theta = 44^o$, new reflection minimas can be observed in the reflection spectrum for the TM polarized wave in Figs. 25(d). The reflection spectrum in Fig. 25(d) is in good agreement with the dispersion relation calculated in Fig. 25(b). Therefore, the continuum of high-order modes is fulled verified and the related results may be applied to a series of practical microcavity applications, such as the switching and fiters [103].

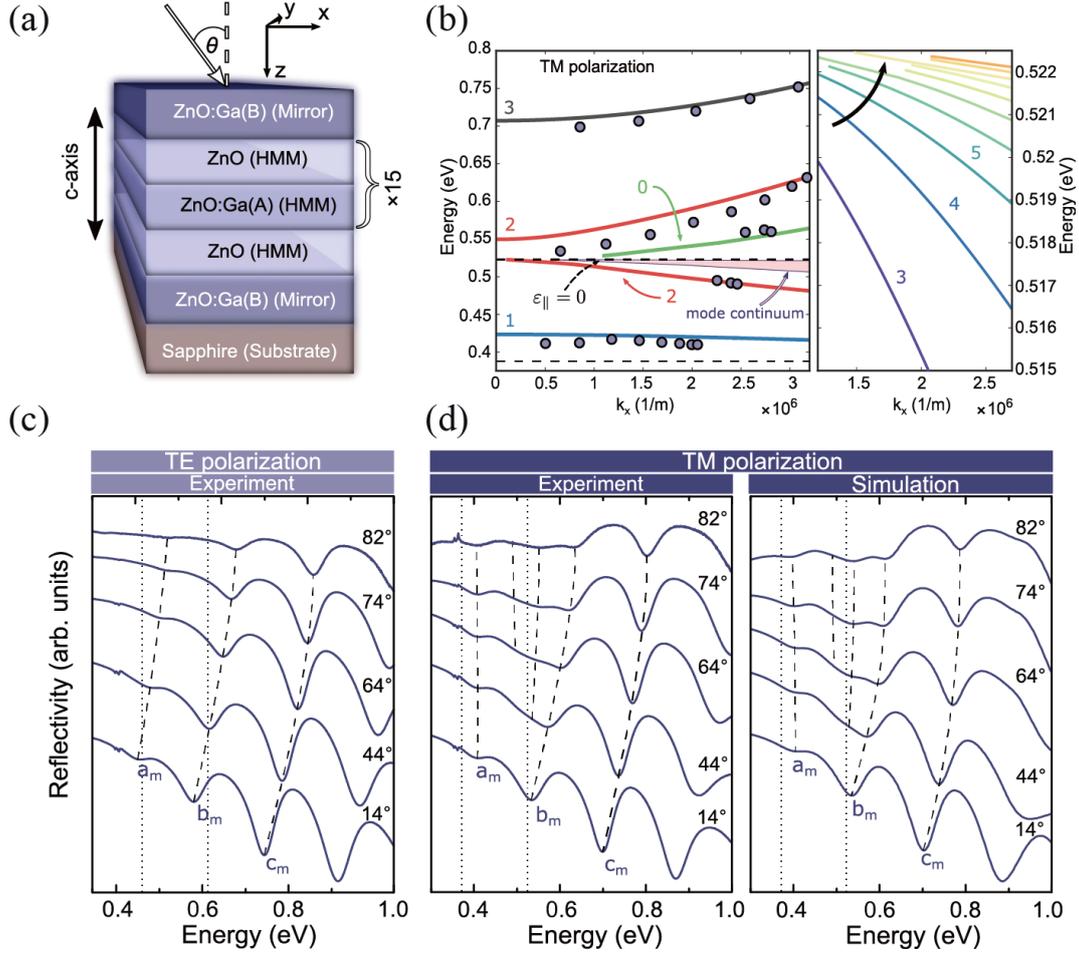

FIG. 25. **Observation of continuum of high-order modes in an optical microcavity with HMM core.** (a) Sketch of the 1D optical microcavity with HMM core, where HMM is realized based on the metal/dielectric multilayer structure. ZnO and heavily doped ZnO: Ga as the dielectric and metal, respectively. (b) Dispersion relation of the microcavity for TM polarization. Inset shows the continuum of high-order modes. The numbers denote the mode order. (c) Measured reflection of the microcavity for TE polarization. (d) Measured and simulated reflection of the microcavity for TM polarization. Reproduced with permission from Travkin et al., Phys. Rev. B 97, 195133 (2018). Copyright 2018, American Physical Society.

## C. Applications

### 1. Lasers

One the most important applications of the optical cavity is the lasers. However, it is difficult to miniaturize the conventional cavity, mainly because the FP resonance condition determines that the size of the cavity is comparable to the wavelength. Although the localized surface plasmon resonance of metal can be used for the small volume localization, the intrinsic loss of metal has a great influence on the $Q$ value of the cavity. The high-$k$ modes of HMM provide a new way to surpass these limitations and realize the subwavelength optical cavity with low-loss, which make the HMM cavities attractive candidates for ultra-small low-threshold lasers [99, 283, 284]. The low threshold spaser based on deep-subwavelength spherical hyperbolic cavities is shown in Fig. 26. Figure 26(a) shows the schematic of the hyperbolic cavity composed of alternating metal and dielectric layers. Take a high-order cavity mode ($l$=4) for example, the simulated field distributions of effective medium and multilayer structure are shown in Fig. 26(b) and 26(c), respectively. It is seen that the metal/dielectric multilayers can be well used in the hyperbolic cavity structure [94]. For a spherical hyperbolic cavity with 7 alternating layers of silver and dielectric, the extinction efficiency ($Q_{ext}$) spectrum is shown in Fig. 26(d) [99]. The first three electric terms ($a_1$, $a_2$, $a_3$) and first magnetic term ($b_1$) of the resonant mode are presented. The field distributions of cavity modes WGM$_{1,1}$ and WGM$_{1,2}$ are shown in Figs. 26(e) and 26(f), respectively. Especially, it is seen that the electric fields with different mode orders are confined within different dielectric shell layers. Therefore, in order to generate the laser output, the active materials with optical gain should be placed in the layer with strong field localization. For cavity mode WGM$_{1,1}$ with the gain medium in the first dielectric layer (painted red for see in the bottom inset), the calculated scattering ($Q_{sca}$) and absorption ($Q_{abs}$) efficiencies as a function of the optical gain $\alpha$ are shown in Fig. 26(g). When $\alpha = 811 \, \text{cm}^{-1}$, the magnitudes of $Q_{sca}$ and $Q_{abs}$ can reach the maximum values, which indicates the lasing threshold of the cavity mode WGM$_{1,1}$. In addition, figure 26(h) shows the linewidth and resonant resonance wavelengths of WGM$_{1,1}$ under different optical gains. It can be clearly seen that the linewidth of cavity mode will greatly decrease near the lasing threshold. Moreover, similar to Fig. 26(g), the $Q_{sca}$ and $Q_{abs}$ efficiencies of cavity mode WGM$_{1,2}$ as a function of the optical gain $\alpha$ are shown in Fig. 26(i). In this case, the the gain medium in the second dielectric layer, which is also painted red for see. From Fig. 26(i), it is seen that the lasing threshold of the cavity mode

WGM$_{1,1}$ corresponds to $\alpha = 1749$ cm$^{-1}$, and the $Q_{sca}$ and $Q_{abs}$ efficiencies reach the highest magnitudes. Therefore, when the gain medium is simultaneously introduced into the first and second dielectric layers, the spaser with dual-wavelength laser output can be realized based on the hyperbolic cavity [99].

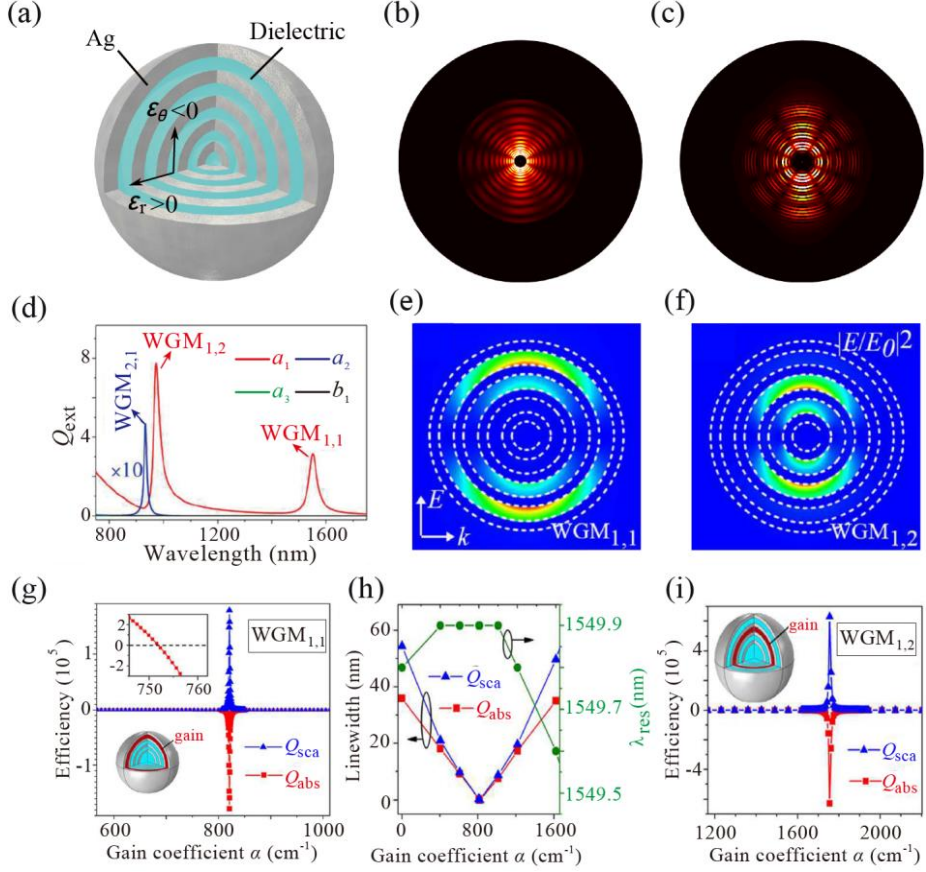

FIG. 26. **Spaser based on deep-subwavelength spherical hyperbolic cavity**. (a) Schematic of the hyperbolic cavity composed of alternating metal and dielectric layers. A WGM supported by (b) a homogeneous hyperbolic cavity and (c) a multilayer structure with mode order $l= 4$. Reproduced with permission from Travkin et al., Phys. Rev. X 4, 021015 (2014). Copyright 2014Author(s), licensed under a Creative Commons Attribution 3.0 License. (d) Extinction efficiency ($Q_{ext}$) of the HMM cavity, which is decomposed into the contributions from the first three electric terms ($a_1$, $a_2$, $a_3$) and the first magnetic term ($b_1$) of Mie expansion. (e) (f) Electric field intensity enhancement distributions of the WGM$_{1,1}$ and WGM$_{1,2}$, respectively. (g) Scattering ($Q_{sca}$) and absorption ($Q_{abs}$) efficiencies calculated at the resonance WGM$_{1,1}$ for the hyperbolic cavities with the same structural parameters but different optical gains. (h) The linewidth and the resonance wavelengths as a function of the optical gain. (i) Same as (g) but calculated at the resonance WGM$_{1,2}$. The gain layer is painted by red for see. Reproduced with permission from Wan et al., Appl. Phys. Lett. 110, 031103 (2017). Copyright 2017, AIP Publishing.

In addition to the gain efficiency, the radiation feedback is also an important factor in the design of miniaturized lasing devices. Recently, Shen et al., experimentally demonstrated that hyperbolic cavity can be used to construct deep-ultraviolet (DUV) plasmonic nanolaser without a long structure, which paves the way for the design of subwavelength nanolaser devices [100]. The DUV nanolaser is realized by placing a hyperbolic cavity on a multiple quantum-well (MQW), and the corresponding schematic is shown in Fig. 27(a). Importantly, the designed hyperbolic cavity merges plasmon resonant modes within the cube and provides a unique resonant radiation feedback to the MQW [100]. Figure 27(b) shows time evolution of the resonant (3,1) mode, and the radiation flows (marked by the arrows) between hyperbolic cavity and MQW is presented in the bottom inset. It is seen that the resonant mode provides a signifcant radiation feld at the position of MQW and provides the needed feedback to MQW. The emission property of the hyperbolic nanolaser is shown in Fig. 27(c). From the measured time resolved photoluminescent (TRPL), the spontaneous decay rate enhancement of hyperbolic cavity can be observed in Fig. 27(c). Compared with the bare MQW sample, the PL of hyperbolic cavity-MQW sample and HMM-MQW sample is enhanced 33 and 1.5 times, respectively. Therefore, the hyperbolic cavity can be used to design low threshold lasers. Especially, even when the HMM-MQW sample showed only spontaneous emission, the hyperbolic cavity-MQW sample will show the lasing in the same pump power range [100]. Moreover, figure 27(d) shows the PL spectrum versus pump power of hyperbolic cavity-MQW sample and the PL peak intensity versus pump power is shown in the inset. It is seen that a sharp lasing peak at 289 nm emerges from a broad spontaneous emission spectrum at a pump power around 106 kW cm$^{-2}$, which corresponses the lasing threshold of the hyperbolic nanolaser. Therefore, based on the high efficiency radiation feedback of hyperbolic cavity mode, the plasmonic nanolaser using a hyperbolic metacavity on an MQW sample provides a new way to enhance the light-matter interaction. The results in Fig. 27 fully verified that the hyperbolic cavities can be employed for nanolaser devices with low lasing threshold at a subwavelength scale.

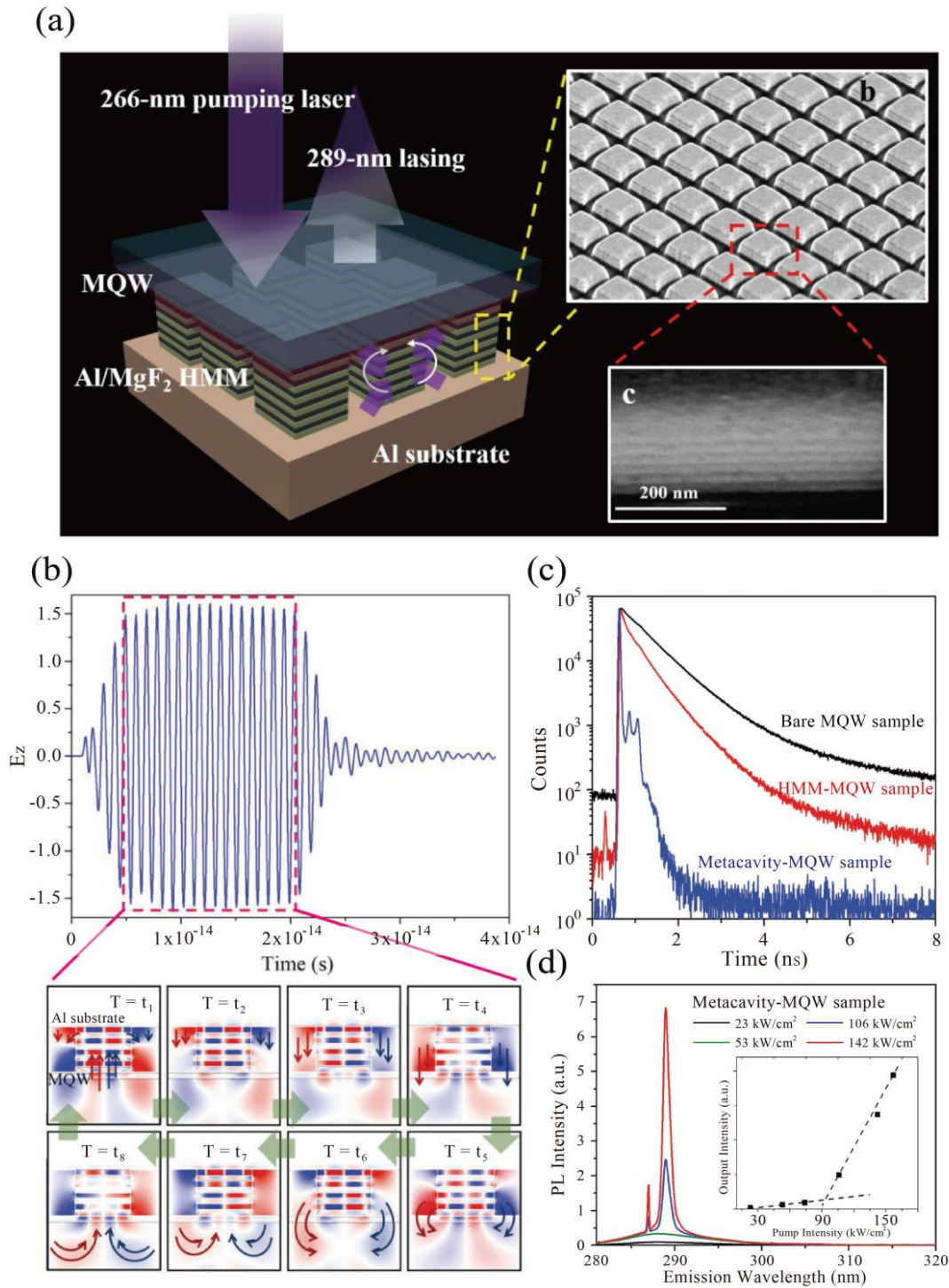

FIG. 27. **DUV plasmonic nanolaser**. (a) Schematic diagram of the DUV plasmonic nanolaser showing an array of square hyperbolic cavity mesas on top of an MQW sample. (b) Time evolution of the resonant (3,1) mode. (c) Time-resolved PL measured as a function of spontaneous emission time for the bare MQW sample, HMM-MQW sample, and hyperbolic cavity-MQW sample. (d) PL spectrum versus pump power of hyperbolic cavity-MQW sample. A narrow lasing peak emerges from a broad emission spectrum. Inset shows the PL peak intensity versus pump power showing the lasing threshold. Reproduced with permission from Shen et al., Adv. Mater. 30, 1706918 (2018). Copyright 2018 Wiley-VCH.

### *2. All-angle filters*

The exceptional dependence of propagating wavevector on polarization, frequency as well as on the angle of incident light in hypercrystals have been systemically introduced [285-291]. Especially, the dispersionless bandgap of 1D hypercrystal has been proposed based on the condition of phase variation compensation [286]. One important application of the dispersionless gap is to design the dispersionless cavity mode. Figure 28(a) shows a hypercrystal with a defect layer in the center of the structure. The hypercrystal consists of alternating thin layers of HMM and dielectric, and the effective HMM is realized by the metal/dielectric multilayers, which is shown in the inset of Fig. 28(a). For the hypercrystal $[(CD)_4B]_3B[(CD)_4B]_3$, in which B and C denote the dielectric and D is the metal, the calculated transmittance as a function of the incident angle and frequency is shown in Fig. 28(b). It is seen that the cavity mode inside the dispersionless gap remains nearly invariant with incident angles, which indicates the dispersionless cavity mode is realized in the 1D hypercrystal. Especially, the transmittance spectra of the structure at three representative angles (0 deg, 30 deg and 60 deg) is shown in Fig. 28(c). It is seen that the positions of cavity modes at different incident angles are nearly unchanged, which is in accordance with Fig. 28(b). In addition, the $Q$ factor of cavity modes nearly unchanged at three incident angles. The dispersionless cavity modes of hyperbolic cavity will possess significant applications for all-angle filters. Moreover, considering the semiconductor HMM (tuned dynamically through optical pumping on a picosecond scale) or the graphene-based HMM (tuned dynamically through an external voltage), the working frequency of all-angle filter can be flexibly tuned [293-300]. The dispersionless cavity mode is also very useful in nonlinear wave mixing and phase-matched [301, 302]. For the traditional cavity mode, phase matching plays an important role in the coherent nonlinear optical process, which can not meet the requirements of all the incident angles [301]. By contrast, the dispersionless cavity mode may provide all-angle phase matching in coherent nonlinear optical process.

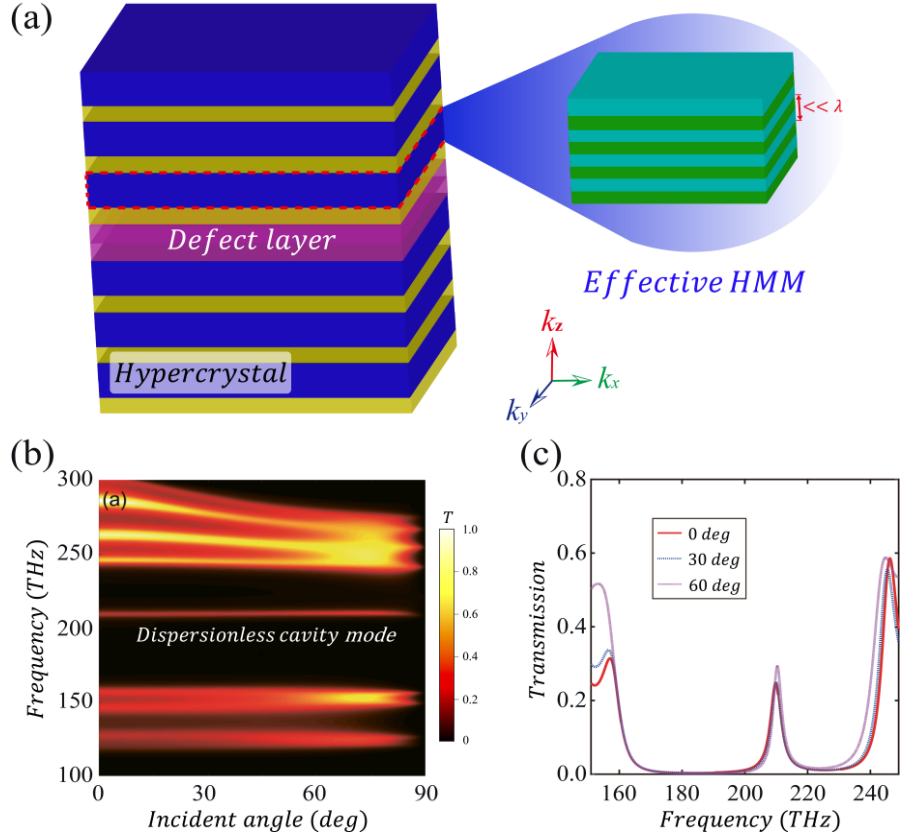

FIG. 28. **Filter based on the defect modes in the hypercrystal**. (a) Schematic of a 1D hypercrystal with a defect layer at the center of the structure. Inset shows that the HMM layer can be mimicked by a metal/dielectric layered structure. (b) Transmittance of the hypercrystal [(CD)₄B]₃B[(CD)₄B]₃ at all incidence angles for TM polarization. (c) Transmittance of the hypercrystal [(CD)₄B]₃B[(CD)₄B]₃ at incident angles of 0°, 30°, and 60°, respectively. 

### 3. Wide-angle biosensors

The edge states in the 1D heterstructure composed of the metal layer and PCs have the properties of localization and polarization-dependence, which can be used to design ultrasensitive optical sensors based on the singularity of the ellipsometric phase (i.e., reflection phase difference between two orthogonal polarizations $\Delta = \varphi_{TM} - \varphi_{TE}$ ) [303, 304]. The underlying physical mechanism is that the frequency of edge modes of TE and TM polarized waves is not overlapped, thus the ellipsometric phase will changed significantly and it can be used to increase the sensitivity of the sensors. However, the bandgap of tranditional 1D all-dielectric PCs for both TE and TM polarized wavers is blueshift along with the increase of the incident angle, thus the edge states of two polarized waves will blueshit. For the case of small angle incidence, the reflection phase difference of two polarized waves is not obvious, which can not achieve high sensitivity. Therefore, for the traditional 1D PC, the incidence angle range of high sensitivity sensor is small. In the integrated optical system, the incident light often has a certain beam width and angular divergence, which will lead to the further reduction of sensing sensitivity. The special dispersion of hyperbolic cavity mode can be used to realized the red shift band gap of TM polarization and the blue shift band gap of TE polarization in 1D PC with HMMs [265]. The schematic of the heterostructure composed of a metal layer and a 1D hypercrystal (in which the HMM layer is realized by the metal/dielectric multilayers) is shown in Fig. 29(a). The reflectance of hypercrystal as a function of incident angle and wavelength is shown in Fig. 29(b). It is seen that the hyperbolic cavity mode in the hypercrystal can realize the red shift in TM polarization and blue shift in TE polarization, thus greatly broadening the working angle range of high sensitivity biosensor. Considering a thick liquid cell filled with bio-solution, whose refractive index is $n_{\text{Bio}}$, the performance of the biosensor based on the hyperbolic cavity at a small incident angle $\theta = 20^o$ is shown in Figs. 29(c) and 29(d). Compared with the results of $nBio = 1.33$ and $nBio = 1.34$ in Fig. 29(c), it can be seen that $|\delta\Delta| = 0.87$ deg for the edge state $\lambda = 1438$ nm. In addition, this biosensor based on the hyperbolic cavity has near-linear response in a relative wide range of refractive index, as shown in Fig. 29(d). Besides, the sensitivity of the proposed biosensor also has been presented in Fig. 29(d), indicating that the proposed biosensor can work with a high refractive index resolution even at a small incident angle. Therefore, hyperbolic cavity realized by the hypercrystal enable the new designs of biosensors with wide-angle and ultrasensitive properties [265].

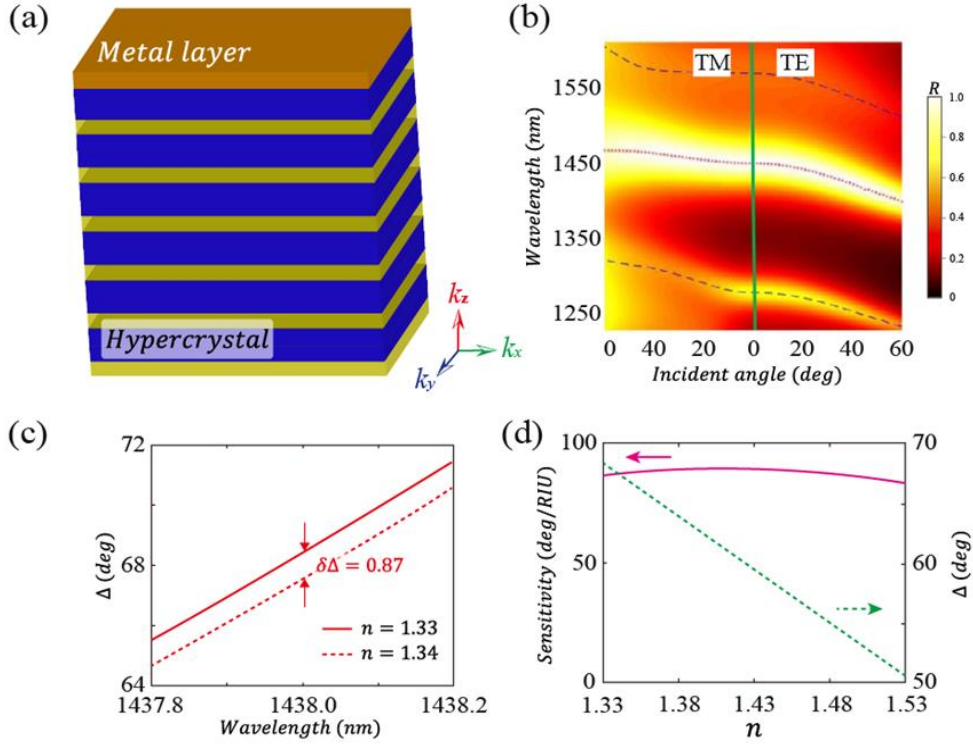

FIG. 29. **Wide-angle ultrasensitive sensor based on the edge modes in the hypercrystal**. (a) Schematic of the heterostructure composed of a metal layer and a 1D PC with HMMs. (b) Reflectance spectra of $[(CD)_2B]_9$ versus incident angle for TM and TE polarizations. Blue dashed lines represent the gap-edges. (c) Ellipsometric phase spectra for $n_{Bio}$ = 1.33 and $n_{Bio}$ = 1.34 at $\theta$ = 20º. (d) Ellipometric phase (green dashed line) and sensitivity (pink solid line) as a function of refractive index at $\theta$ = 20º and $\lambda$ = 1438 nm. Reproduced with permission from Wu et al., Opt. Express 27, 24835 (2019). Copyright 2019, OSA Publishing.

## IV. CONCLUSIONS AND OUTLOOK

The optical cavity with metamaterial (i.e. metacavity) is of great significance in fundamental and applied physics. On the one hand, the optical cavity can confine the EM wave in a small scale, which can effectively improve the photon DOS thus enhancing the interaction between light and the matter. On the other hand, metamaterials can control the transmission, radiation and coupling of EM wave arbitrarily, so it can realize many novel physical phenomena which are difficult to realize in traditional optical cavities, and can be used to construct more functional optical devices.

In this review, two kinds of special metacavities, zero-index metacavity and hyperbolic metacavity, are systematically introduced. (1) For the zero-index cavity, it can achieve novel geometry (shape, topology, size) independent cavity mode, uniformed field enhancement, and directional radiation. Especially for the matched EZI heterostructure structure, it provides a good solution for the construction of subwavelength optical cavity, and has been proved to significantly enhance the interaction between light and matter, such as nonlinear enhancement, magneto-optical effect enhancement and so on. In addition, it is interesting that the heterostructure composed of two kinds of topological distinguished 1D PCs with symmetrical configuration can realize the topologically protected edge state, which has been confirmed from microwave to ultra- X-ray band. This is of great significance in the realization of optical devices sensitive to structural errors. Moreover, the applications of zero-index metacavity in switching, nonreciprocal transmission and collective coupling are introduced in detail. (2) For the hyperbolic metacavity, it has been observed that it has anisotropic resonance characteristics, and can observe the anomalous scaling law, scale independent cavity modes and continuous cavity modes. In particular, the special dispersion of hyperbolic metacavity can be used to realize miniaturization of low-loss laser. And the special mode distribution of hyperbolic cavity mode can be used to enhance the radiation feedback and reduce the lasing threshold. In addition, for the hypercrystal constructed by HMMs, by adjusting the structure parameters, the dispersionless cavity mode can be realized and used for all-angle filters. Similarly, based on the dispersion control, hypercrystal can also be used to design highly sensitive biosensors. In general, the zero-index and hyperbolic metacavities provide new and efficient means for EM wave control, and provide new ways for the design of novel optical devices, which are expected to be applied in the future of photonic integration.

So far, ZIMs and HMMs have been widely used to study the novel transmission properties of EM waves [305-310], including absorption [311-313], filtering [314, 315], scattering [316-318], splitting [319, 320], guide modes [321-325], surface waves [326-328], high-harmonic waves [329, 330] and topological modes [331-333]. The related physical properties in metacavities need to be further studied. Especially, a new type of metamaterial, LCMM, has been proposed [281, 334-337]. As a special kind of anisotropic material, it has the characteristics of both ZIM and HMM. Therefore, the performance of this new metamaterial for optical cavity design needs to be explored in the future. In addition, biaxial metacavity is also a very interesting research topic [338-341]. Very recently, the realization of low-loss metamaterials with natural 2D materials greatly promotes the development of plane EM wave control [342-345]. Importantly, with the development of twist-optics [346-348], interlayer rotation of 2D materials/metasurfaces provides a new way to study the topological phase transition of IFC [349-360]. Therefore, the interlayer coupling effect of planar optical metacavity (including zero-index metacavity and hyperbolic metacavity) and its novel functions will also provide ways for the design of next generation optical cavity devices. In general, the metacavity can flexibly control the near-field of EM wave, and it is expected to be applied in wireless power transfer [361-365], wireless sensing [366-370], magnetic resonance imaging [371-376] and so on. In particular, although the current researches on ZIMs, HMMs, and the related metacavities are mainly focused on EM wave system, the related research results are expected to be well extended to acoustic [377-380] and thermal [381-387] systems, and it is to be expected that more novel devices based on metacavities will be developed in the future.

## ACKNOWLEDGMENTS

This work was supported by the National Key R&D Program of China (Grant No. 2016YFA0301101), the National Natural Science Foundation of China (NSFC) (Grant Nos. 12004284, 11774261, and 61621001), the Shanghai Science and Technology Committee (Grant No. 18JC1410900), the Shanghai Super Postdoctoral Incentive Program, and the China Postdoctoral Science Foundation (Grant Nos. 2019TQ0232 and 2019M661605).